\documentclass[11pt]{article}

\usepackage{fullpage}
\usepackage{amsmath,amssymb,amsfonts,enumerate,bbm}
\usepackage{caption}
\usepackage{url}
\usepackage{ifthen}
\usepackage{ifpdf}
\ifpdf
\usepackage[pdftex]{graphicx}
\usepackage[pdftex]{hyperref}
\else
\usepackage[dvips]{graphicx}
\fi

\usepackage{subfigure}

\usepackage{color,soul}

\newcommand{\Comment}[1]{}

\usepackage{times}

\long\def\commabs #1\commabsend{}
\long\def\commful #1\commfulend{#1}

\def\inline#1:{\par\vskip 7pt\noindent{\bf #1:}\hskip 10pt}
\def\Proof{\par\noindent{\bf Proof:~}}
\def\blackslug{\hbox{\hskip 1pt \vrule width 4pt height 8pt
    depth 1.5pt \hskip 1pt}}

\newtheorem{theorem}{Theorem}[section]

\newtheorem{corollary}[theorem]{Corollary}

\newtheorem{claim}[theorem]{Claim}
\newtheorem{fact}[theorem]{Fact}
\newtheorem{definition}[theorem]{Definition}

\def\inline#1:{\par\vskip 7pt\noindent{\bf #1:}\hskip 10pt}
\def\Proof{\par\noindent{\bf Proof:~}}
\def\blackslug{\hbox{\hskip 1pt \vrule width 4pt height 8pt
    depth 1.5pt \hskip 1pt}}
\def\QED{\quad\blackslug\lower 8.5pt\null\par}

\def\cC{{\cal C}}
\def\cA{{\cal A}}
\def\cE{{\cal E}}

\def\cL{{\cal L}}

\def\cN{{\cal N}}
\def\cI{{\cal I}}

\def\cY{{\cal Y}}

\newcommand{\LD}{\mbox{\rm LD}}
\newcommand{\NLD}{\mbox{\rm NLD}}

\newcommand{\BPLD}{\mbox{\rm BPLD}}

\newcommand{\tree}{\mbox{\tt Tree}}

\newcommand{\leader}{\mbox{\tt AMOS}}
\newcommand{\leaderk}{\mbox{\tt AMOS-$k$}}
\newcommand{\leadera}{\mbox{\tt AMOS-$a$}}

\newcommand{\Alphabet}{\Sigma}

\def\NewSymb{\otimes}

\newcommand{\SecuritySize}{\mbox{$\ell$}}

\def\distalg{{\cA}}
\newcommand{\local}{{\cal LOCAL}}
\newcommand{\congest}{{\cal CONGEST}}
\newcommand{\inp}{\mbox{\rm\bf x}}

\newcommand{\out}{\mbox{\rm out}}
\newcommand{\id}{\mbox{\rm Id}}

\newcommand{\yes}{\mbox{``yes''}}
\newcommand{\no}{\mbox{``no''}}

\long\def\jump#1\finjump{}

\usepackage{algorithmicx}
\usepackage{algorithm}
\usepackage{eepic}
\usepackage[noend]{algpseudocode}

\newcounter{smallitemizec}
\newenvironment{smallitemize}
{   \setcounter{smallitemizec}{0}
  \begin{list}{$\bullet$}
    {\usecounter{smallitemizec}
      \setlength{\parsep}{0pt}
      \setlength{\itemsep}{0pt}}
    }{ \end{list} }

\begin{document}

\pagenumbering{arabic}

\title{\bf Randomized Distributed Decision}
\author{
Pierre Fraigniaud \hbox{\hskip 40pt} Amos Korman \\
CNRS and University Paris Diderot, France.
{\small\sf \{pierre.fraigniaud,amos.korman\}@liafa.jussieu.fr}
\thanks{Supported by the ANR projects DISPLEXITY and PROSE, and by the INRIA project GANG.}
\\
\and Merav Parter \hbox{\hskip 40pt}  David Peleg \\
The Weizmann Institute of Science, Rehovot, Israel.
{\small\sf \{merav.parter,david.peleg\}@weizmann.ac.il}
\thanks{Supported in part by the Israel Science Foundation (grant 894/09),
the US-Israel Binational Science Foundation (grant 2008348),
the Israel Ministry of Science and Technology (infrastructures grant),
and the Citi Foundation.}
}

\date{}

\maketitle

\begin{abstract}
The paper tackles the power of randomization in the context of locality by analyzing the ability to ``boost'' the success probability of deciding a distributed language. The main outcome of this analysis is that the distributed computing setting contrasts  significantly with the sequential one as far as randomization is concerned. Indeed, we prove that in some cases, the ability to increase the success probability for deciding  distributed languages is rather limited.

Informally,  a \emph{$(p,q)$-decider} for a language $\cL$ is a distributed randomized algorithm which {\em accepts} instances in $\cL$ with probability at least $p$ and {\em rejects} instances outside of $\cL$ with probability at least $q$.  It is known that every hereditary language
that can be decided  in $t$ rounds by a  $(p,q)$-decider, where $p^2+q>1$,
  can actually  be decided \emph{deterministically} in $O(t)$ rounds.
In one of our results we give evidence supporting the conjecture that the above statement holds for all distributed languages and not only for hereditary ones. This is achieved  by considering the restricted case of path topologies.

We then turn our attention to the range below the aforementioned threshold, namely, the case where $p^2+q\leq1$.  For $k\in \mathbb{N}^*\cup \{\infty\}$, we define  the class $B_k(t)$ to be the set of all languages decidable in at most $t$ rounds by a $(p,q)$-decider, where $p^{1+\frac{1}{k}}+q>1$.
It is easy to see that every language is decidable  (in zero rounds) by a $(p,q)$-decider satisfying $p+q=1$. Hence, the hierarchy $B_k$ provides a spectrum of complexity classes between determinism ($k=1$, under the above conjecture) and complete  randomization ($k=\infty$). We prove that all these classes are separated, in a strong sense: for every integer $k\geq 1$, there exists a language $\cL$ satisfying $\cL\in B_{k+1}(0)$ but $\cL\notin B_k(t)$ for any $t=o(n)$. In addition, we show that $B_\infty(t)$ does not contain all languages, for any $t=o(n)$. In other words, we obtain the following hierarchy:
\commful
$$B_1(t)\subset B_2 (t)\subset \cdots \subset B_\infty(t) \subset \mbox{All}~.$$
\commfulend
\commabs
$$B_1(t)\subset B_2 (t)\subset \cdots
\subset B_k(t)\subset \cdots \subset B_\infty(t) \subset \mbox{All}~.$$
\commabsend
Finally, we show that if the inputs can be restricted in certain ways, then the ability to boost the success probability becomes almost null, and, in particular, derandomization is not possible even beyond the threshold $p^2+q=1$.

All our results hold with respect to the $\local$ model of computation  as well as with respect to  the  $\congest(B)$ model, for $B=O(1)$.

\end{abstract}


\section{Introduction}

\subsection{Background and Motivation}

The impact of randomization on computation is one of the most central questions in  computer science. In particular, in the context of distributed computing, the question of whether randomization helps in improving locality for construction problems has been studied extensively.
While most of these studies were problem-specific,
several attempts have been made for tackling this question from a more  general and unified perspective. For example, Naor and Stockmeyer~\cite{NS93} focus on a class of problems called LCL (essentially a subclass of the class $\LD$ discussed below), and show that if there exists a randomized algorithm that constructs a solution for a problem in LCL in a constant number of rounds, then there is also a constant time deterministic algorithm constructing a solution for that problem.

Recently, this question has been studied in the context of {\em local decision}, where one aims at deciding locally whether a given global input instance belongs to some specified language~\cite{FKP11}. The localities of deterministic algorithms and randomized Monte Carlo algorithms are compared in~\cite{FKP11}, in the $\local$ model (cf.~\cite{PelB00}). One of the main results of~\cite{FKP11} is that randomization does not help for locally deciding {\em hereditary} languages if the success probability is beyond a certain guarantee threshold.
More specifically, a \emph{$(p,q)$-decider} for a language $\cL$ is a distributed randomized Monte Carlo algorithm that {\em accepts} instances in $\cL$ with probability at least $p$ and {\em rejects} instances outside of $\cL$ with probability at least $q$.  It was shown in~\cite{FKP11} that every hereditary language
that can be decided  in $t$ rounds by a  $(p,q)$-decider, where $p^2+q>1$,
  can actually  be decided \emph{deterministically} in $O(t)$ rounds. On the other hand, \cite{FKP11} showed that the aforementioned threshold is sharp, at least when hereditary languages are concerned.
In particular, for every $p$ and $q$,  where $p^2+q\leq1$, there exists an heredirtary language that cannot be decided deterministically in $o(n)$ rounds, but can be decided in zero number of rounds by a $(p,q)$-decider.

 In one of our results we provide  evidence supporting the conjecture that the above statement holds for all distributed languages and not only for hereditary ones. This is achieved  by considering the restricted case of path topologies.
In addition, we present a more refined analysis for the family of languages
that can be decided randomly but not deterministically. That is, we focus
on the family of languages that can be decided locally by a $(p,q)$-decider, where $p^2+q\leq1$, and introduce an infinite hierarchy of classes within this family, characterized by the specific relationships between the parameters $p$ and $q$. As we shall see, our results imply  that the distributed computing setting contrasts  significantly with the sequential one as far as randomization is concerned. Indeed,  we prove that in some cases, the ability to increase the success probability for deciding  distributed languages is very limited.

\subsection{Model}\label{subsec:model}

We consider the $\local$ model (cf.~\cite{PelB00}), which is a standard distributed computing model capturing the essence of spatial locality. In this model, processors are woken up simultaneously, and computation proceeds in fault-free synchronous rounds during which
every processor exchanges messages of unlimited size with its neighbors, and performs arbitrary computations on its data.
It is important to stress that  all the  algorithmic constructions that we employ in our positive results use messages of constant size
 (some of which do not use any communication at all).
Hence, all our results apply not only to the $\local$ model of computation but also to more restricted models, for example,
 the $\congest(B)$ model\footnote{Essentially, the $\congest(B)$   model is similar to the $\local$ model, except that the message size is assumed to be bounded by at most $B$ bits (for more details, see~\cite{PelB00}).},  where $B=O(1)$.

 A distributed algorithm $\distalg$ that runs on a graph $G$ operates separately on each connected component of $G$, and nodes of a component $C$ of $G$ cannot distinguish the underlying graph $G$ from $C$. For this reason, we consider  connected graphs only.

We focus on \emph{distributed decision tasks}. Such a task is characterized by a finite or infinite set $\Sigma$ of symbols (e.g., $\Sigma=\{0,1\}$, or $\Sigma=\{0,1\}^*$), and by a {\em distributed language} $\cL$ defined on this set of symbols (see below). An {\em instance} of a distributed decision task is a pair $(G,\inp)$ where $G$ is an $n$-node connected graph, and $\inp\in\Sigma^n$, that is, every node $v\in V(G)$ is assigned as its {\em local input} a value $\inp(v)\in \Sigma$. (In some cases, the local input of every node is empty, i.e., $\Sigma=\{\epsilon\}$, where $\epsilon$ denotes the empty binary string.) We define a {\em distributed language} as a decidable collection $\cL$ of instances\footnote{Note that an undecidable collection of instances remains undecidable in the distributed setting too.}.

In the context of distributed computing, each processor must produce a boolean output, and the decision is defined by the conjunction of the processors���½ outputs, i.e., if the instance belongs to the language, then all processors must output ``yes'', and otherwise, at least one processor must output ``no''.
Formally, for a distributed language $\cL$, we say that a distributed algorithm $\distalg$ \emph{decides} $\cL$ if and only if for every instance $(G,\inp)$ and id-assignment $\id$, every node $v$ of $G$ eventually terminates and produces an output denoted $\out_\distalg(G,\inp,\id,v)$, which is either \yes\/ or \no, satisfying the following decision rules:
\begin{smallitemize}
\item If $(G,\inp)\in \cL$ then  $\out_\distalg(G,\inp,\id,v)=\yes$ for every node $v\in V(G)$~;
\item If $(G,\inp)\notin \cL$ then  $\out_\distalg(G,\inp,\id,v)=\no$ for at least one node $v\in V(G)$~.
\end{smallitemize}

Observe that decision problems provide a natural framework for tackling fault-tolerance: the processors have to collectively check whether the network is fault-free, and a node detecting a fault raises an alarm. In fact, many natural problems can be phrased as decision problems, for example:   ``is the network planar?" or ``is there a unique leader in the network?''. Moreover, decision problems occur naturally when one aims at checking the validity of the output of a computational task, such as ``is the produced coloring legal?", or ``is the constructed subgraph an MST?".

The class of decision problems that can be solved in at most $t$ communication rounds is denoted by $\LD(t)$, for \emph{local decision}. More precisely, let $t$ be a  function of triplets $(G,\inp,\id)$, where $\id$ denotes the identity assignment to the nodes of $G$. Then $\LD(t)$ is the class of all distributed languages that can be decided by a  distributed algorithm that runs in at most $t$ communication rounds. The randomized (Monte Carlo 2-sided error) version of the class $\LD(t)$ is denoted $\BPLD(t,p,q)$, which stands for \emph{bounded-error probabilistic local decision}, and provides an analog of BPP for distributed computing, where $p$ and $q$ respectively denote the yes-error and the no-error guarantees. More precisely, a {\em randomized} distributed algorithm is a distributed algorithm $\distalg$ that enables every node $v$, at any round $r$ during its execution, to generate a certain number of random bits. For constants $p,q\in(0,1]$, we say that a randomized distributed algorithm $\distalg$ is a {\em $(p,q)$-decider} for $\cL$, or, that it decides $\cL$ with \yes\/ success  probability $p$ and \no\/ success  probability $q$, if and only if for every instance $(G,\inp)$ and id-assignment $\id$, every node of $G$ eventually terminates and outputs \yes\/ or \no, and the following properties are satisfied:
\begin{smallitemize}
\item If $(G,\inp)\in \cL$ then $\Pr[\forall v\in V(G)$, \\
\hbox{\hskip 40pt} $\out_\distalg(G,\inp,\id,v)=\mbox{\yes }] \ge p$~;
\item If $(G,\inp)\notin \cL$ then $\Pr[\exists  v\in V(G)$, \\
\hbox{\hskip 40pt} $\out_\distalg(G,\inp,\id,v)=\mbox{\no}] \ge q$~.
\end{smallitemize}

The probabilities in the above definition are taken over all possible coin tosses performed by the nodes. The running time of a $(p,q)$-decider executed on a node $v$ depends on the triple $(G,\inp,\id)$ and on the results of the coin tosses. In the context of a randomized algorithm, $T_v(G,\inp,\id)$ denotes the maximal running time of the algorithm on $v$ over all possible coin tosses, for the instance $(G,\inp)$ and id-assignment $\id$. Now, just as in the deterministic case, the running time $T$ of the $(p,q)$-decider is the maximum running time over all nodes. Note that by definition of the distributed Monte-Carlo algorithm, both $T_v$ and $T$ are deterministic. For constant $p,q\in(0,1]$ and a function $t$  of triplets $(G,\inp,\id)$, $\BPLD(t,p,q)$ is the class of all distributed languages that have a randomized distributed $(p,q)$-decider running in time at most~$t$ (i.e., can be decided in time at most $t$ by a randomized distributed algorithm with \yes\/ success  probability $p$ and \no success probability $q$).

Our main interest within this context is in studying the connections between the classes $\BPLD(t,p,q)$. In particular, we are interested in the question of whether one can ``boost'' the success probabilities of a $(p,q)$-decider. (Recall that in the {\em sequential} Monte Carlo setting, such ``boosting'' can easily be achieved by repeating the execution of the algorithm a large number of times.) Our starting point is the recent result of~\cite{FKP11} that, for the class of hereditary languages (i.e., closed under sub-graphs),
the relation $p^2+q=1$ is a sharp threshold for randomization. That is,
for hereditary languages, $\bigcup_{p^2+q>1}\BPLD(t,p,q)$ collapses to $\LD(O(t))$, but for any $p,q\in(0,1]$ such that $p^2+q\leq1$ there exists a language $\cL\in  \BPLD(0,p,q)$, while $\cL\notin  \LD(t)$ for any $t=o(n)$. We conjecture that the hereditary assumption can be removed and we give some evidence supporting this conjecture. Aiming at analyzing the collection of classes
$\bigcup_{p^2+q\leq1}\BPLD(t,p,q)$, we consider the following set of classes:
\[ B_{k}(t)= \bigcup_{p^{1+1/k}+q>1} \BPLD(t,p,q)\]
for any positive integer $k$, as well as the class
\[ B_{\infty}(t)= \bigcup_{p+q>1} \BPLD(t,p,q)~.\]
Hence, our conjecture states that $B_1(t)=\LD(O(t))$. Note that the class $\bigcup_{p+q \geq 1} \BPLD(0,p,q)$ contains \emph{all} languages, using a $(1,0)$-decider that systematically returns \yes\/ at every node (without any communication). Hence, the classes $B_k$ provide a smooth spectrum of randomized distributed complexity classes, from the class of deterministically decidable languages (under our conjecture) to the class of all languages. The ability of boosting the success probabilities of a $(p,q)$-decider is directly related to the question of whether these classes are different, and to what extent.

\subsection{Our results}

One of the main outcomes of this paper is a proof that boosting success probabilities in the distributed setting appears to be quite limited.  By definition, $B_{k}(t)\subseteq B_{k+1}(t)$ for any $k$ and $t$. We prove that these inclusions are strict. In fact, our separation result is much stronger. We prove that there exists a language in $B_{k+1}(0)$ that is not in  $B_k(t)$ for any $t=o(n)$. Moreover, we prove that $\tree \notin B_\infty(t)$ for any $t=o(n)$, where $\tree =\{(G,\epsilon):\mbox{$G$ is a tree}\}$. Hence, $B_\infty(t)$ does not contain all languages, even for $t=o(n)$. In summary, we obtain the following hierarchy.
\commful
$$B_1(t)\subset B_2 (t)\subset \cdots \subset B_\infty(t) \subset \mbox{All}~.$$
\commfulend
\commabs
$$B_1(t)\subset B_2 (t)\subset \cdots
\subset B_k(t)\subset \cdots \subset B_\infty(t) \subset \mbox{All}~.$$
\commabsend
These results demonstrate that boosting the probability of success might be doable, but only from a $(p,q)$ pair satisfying $p^{1+1/(k+1)}+q>1$ to a $(p,q)$ pair satisfying $p^{1+1/k}+q>1$ (with the extremes excluded). It is an open question whether $B_{k+1}(t)$ actually collapses to  $\BPLD(O(t),p,q)$, where $p^{1+1/k}+q=1$, or whether there exist intermediate classes.

Recall that every hereditary language in $B_{1}(t)$ is also in $\LD(O(t))$~\cite{FKP11}. We conjecture that this derandomization result holds for all languages and not only for hereditary ones. We give evidence supporting this conjecture by showing that restricted to path topologies, finite input and constant running time $t$, the statement $B_{1}(t)\subseteq \LD(O(t))$ holds without assuming the hereditary property. This evidence seems to be quite meaningful especially since
 all our separation results hold even if we restrict ourselves to decision problems on  path topologies.

Finally, we show that the situation changes  drastically if the distribution of inputs can be  restricted in certain ways.
Indeed, we show that for every two reals $0< r < r'$, there exists a language in $C_{r'}(0)$ that is not in  $C_r(t)$ for any $t=o(n)$, where the $C$-classes are the extension of the $B$-classes to decision problems in which the inputs can be restricted.


All our results hold not only with respect to  the $\local$ model  but also with respect to more restrictive models of computation such as the $\congest(B)$ model (for  $B=O(1)$).

\subsection{Related work}
The notion of local decision and local verification of languages has
received quite a lot of attention recently. In the $\cal{LOCAL}$ model,
for example, solving a decision problem requires the processors to
independently inspect their local neighborhood and collectively decide
whether the global instance belongs to some specified language.
Inspired by classical computation complexity theory, Fraigniaud et al.\!\!\!
\cite{FKP11} suggested  that the study of decision problems may lead to new
structural insights also in the more complex distributed computing setting.
Indeed, following that paper, efforts were made to form a fundamental
computational complexity theory for distributed decision problems in
various other aspects of distributed computing
\cite{FKP11,FP12,FRT11,FRT12}.

The classes $\LD$, $\NLD$ and $\BPLD$ defined in \cite{FKP11} are
the distributed analogues of the classes  P, NP and BPP, respectively.
The contribution of \cite{FKP11} is threefold: it establishes
the impact of nondeterminism, randomization, and randomization + nondeterminism,
on local computation. This is done by proving structural results,
developing a notion of local reduction and establishing completeness results.
One of the main results is the existence of a sharp threshold for randomization,
above which randomization does not help (at least for hereditary languages).
More precisely the $\BPLD$ classes were classified into two:
below and above the randomization threshold. The current paper ``zooms''
into the spectrum of classes below the randomization threshold, and
defines a hierarchy of an infinite set of $\BPLD$ classes, each of which
is separated from the class above it in the hierarchy.

The question of whether randomization helps in improving locality
for construction problems has been studied extensively.
Naor and Stockmeyer~\cite{NS93} considered a subclass of $\LD(O(1))$,
called LCL\footnote{LCL is essentially $\LD(O(1))$ restricted to
languages involving graphs of constant maximum degree and processor inputs
taken from a set of constant size.}, and studied the question of how to compute
in $O(1)$ rounds the constructive versions of decision problems in LCL.
The paper demonstrates that randomization does not help, in the sense that
if a problem has a local Monte Carlo randomized algorithm, then it also
has a local deterministic algorithm.
There are several differences between the setting of \cite{NS93} and ours.
First, \cite{NS93} considers the power of randomization for
\emph{constructing} a solution, whereas we study the power of
randomization  for \emph{deciding} languages\footnote{There is a
fundamental difference between such tasks when locality is concerned.
Indeed, whereas the validity of constructing  a problem in LCL is
local (by definition), the validity in our setting is ``global'',
in the sense that in an illegal instance, it is sufficient that at
least one vertex in the entire network outputs \no.}. Second, while
\cite{NS93} deals with constant time computations, our separation results
apply to arbitrary time computations, potentially depending on the size of the
instance (graph and input). To summarize, the different settings imply
different impacts for randomization: while this current paper as well
as  \cite{FKP11} show that randomization can indeed help for improving
locality of decision problems, \cite{NS93} shows that when it comes to constructing a solution for a problem in LCL in constant time, randomization does not help.
\commabs
From a less unified approach,
\commabsend
The question of whether randomization helps
for constructing solutions to some {\em specific} problems in localized computational models such as MIS, $(\Delta+1)$-coloring, and maximal matching has
been also studied in \cite{ABI,BM09,K09,L86,N91,PS96,SW10}.
\commabs
To date, there exists evidence that, for some problems at
least, randomization does not help.
For instance, \cite{N91} proves this for 3-coloring the ring.
In fact, for low degree graphs, the gaps between the efficiencies
of the best known randomized and deterministic algorithms for problems like
MIS, $(\Delta+1)$-coloring, and maximal matching are very small.
On the other hand,
for graphs of arbitrarily large degrees, there seem to be indications
that randomization does help, at least in some cases.
For instance, $(\Delta+1)$-coloring can be randomly computed in expected
$O(\log n)$ communication rounds on $n$-node graphs~\cite{ABI,L86},
whereas the best known deterministic algorithm for this problem
performs in $2^{O(\sqrt{\log n})}$ rounds~\cite{PS96}.
$(\Delta+1)$-coloring algorithms whose performance is expressed also
in terms of the maximum degree $\Delta$ illustrate this phenomenon as well.
Specifically, \cite{SW10} shows that $(\Delta+1)$-coloring can be randomly
computed in expected $O(\log \Delta +\sqrt{\log n})$ communication rounds,
whereas the best known deterministic algorithm
performs in $O(\Delta+\log^* n)$ rounds~\cite{BM09,K09}.
\commabsend
The original theoretical basis for nondeterminism in local
computation was laid by the theories of \emph{proof-labeling schemes}~\cite{GS11, KK07,KKM11,KKP10}, which resemble the notion of NLD, and
\emph{local computation with advice} \cite{DP12,FGIP07,FIP10,FKL07}.
These notions also bear some similarities to the notions of
{\em local detection}~\cite{AKY97},
{\em local checking}~\cite{APV}, or {\em silent stabilization}~\cite{silent},
which were introduced in the context of self-stabilization~\cite{D74}.
In addition, NLD seems to be related also to the theory of
{\em lifts}~\cite{Linial01}.

Finally, the classification of decision problems in distributed
computing has been studied in several other models. For example,
\cite{DHKKNPPW} and \cite{KKP11} study specific decision problems in
the $\cal{CONGEST}$ model.
In addition, decision problems have been studied in the asynchrony discipline
too, specifically in the framework of {\em wait-free computation}
\cite{FRT11,FRT12} and {\em mobile agents  computing} \cite{FP12}.
In the wait-free model, the main issues are not spatial constraints but timing
constraints (asynchronism and faults). The main focus of  \cite{FRT12} is
deterministic protocols aiming at studying the power of the ``decoder'',
i.e., the interpretation of the results. While this paper essentially
considers the AND-checker, (as a global ``yes" corresponds to all processes
saying ``yes"), \cite{FRT12} deals with other interpretations,
including more values (not only ``yes" and ``no"), with the objective of
designing checkers that use the smallest number of values.

\section{Preliminaries}

This section
recalls some previous results from the literature, to be used throughout in the paper.
Let us first
recall that in the $\local$ (respectively $\congest(B)$)  model, processors perform in synchronous rounds, and, in each round, every processor
(1)~sends messages of arbitrary (resp., $O(B)$ bits) size  to its neighbors,
(2)~receives messages from its neighbors, and
(3)~performs arbitrary individual computations.
After a number of rounds (that may depend on the network $G$ connecting the processors, and may vary among the processors,
since nodes have different identities, potentially different inputs, and are typically located at non-isomorphic positions in the network),
every processor $v$ terminates and generates its output.

Consider a distributed $(p,q)$-decider $\distalg$ running in a network $G$ with input $\inp$ and identity assignment $\id$
(assigning distinct integers to the nodes of $G$).
The output of processor $v$ in this scenario  is denoted
by $\out_\distalg(G,\inp,\id,v)$, or simply $\out(v)$ when the parameters are clear from the context. In the case of decision problem, $\out(v)\in\{\yes,\no\}$ for every processor $v$.

An $n$-node path $P$ is represented as a sequence
$P=(1,\ldots,n)$, oriented from left to right. (However, node $i$ does not know its position in the path.)
Given an instance $(P, \inp)$ with ID's $\id$ and a subpath $S \subset P$, let $\inp_{S}$ (respectively  $\id_{S}$) be the restriction of $\inp$ (resp., $\id$) to $S$. We sometimes refer to subpath $S=(u_i, \ldots, u_j) \subset P$ as $S=[i,j]$.
For a set $U\subseteq V(G)$, let $\cE(G,\inp,\id,U)$ denote the event that,
when running $\distalg$ on $(G,\inp)$ with id-assignment $\id$,
all nodes in $U$ output ``yes''. Given a language $\cL$, an instance $(G,\inp)$ is called \emph{legal} iff $(G,\inp) \in \cL$.

Given a time bound $t$, a subpath $S=[i,j]$ is called an \emph{internal} subpath of $P$ if $i\geq t+2$ and $j \leq n-t-1$. Note that if the subpath $S$ is internal to $P$, then when running a $t$-round algorithm, none of the nodes in $S$ ``sees'' the endpoints of $P$.

We now define an important concept, which is crucial in the proofs of our separation results.

\begin{definition}
Let $S$ be a subpath of $P$. For $\delta \in [0,1]$, $S$ is said to be a $(\delta,\lambda)$-\emph{secure} subpath if $|S|\geq \lambda$ and $\Pr[\cE(P,\inp,\id,V(S))] \geq 1-\delta$.
\end{definition}

We typically use $(\delta,\lambda)$-secure subpaths for values of $\lambda\geq 2t+1$ where $t$ is the running time of the $(p,q)$-decider $\distalg$ on $(P,\inp)$ for some fixed identity assignment $\id$. Indeed, it is known~\cite{FKP11} that if $(P,\inp) \in \cL$, then every long enough subpath $S$ of $P$ contains an internal $(\delta,\lambda)$-secure subpath~$S'$. More precisely, define
\begin{equation}
\label{eq:defsecuritysize}
\SecuritySize(\delta,\lambda) = 4(\lambda+2t) \lceil \log p/ \log(1-\delta) \rceil.
\end{equation}
We have the following:
\begin{fact}[\cite{FKP11}]\label{fact-secure}
\label{cl:sec_zone_guarantee}
Let $(P,\inp) \in \cL$, $\delta \in [0,1]$, $\lambda\geq 1$.  Then for every $\SecuritySize(\delta,\lambda)$-length subpath $S$ there is a subpath $S'$ (internal to $S$) that is $(\delta,\lambda)$-secure.
\end{fact}
For completeness, we provide the proof of this fact in the Appendix.
To avoid cumbersome notation, when $\lambda=2t+1$, we may omit it and refer to $(\delta,2t+1)$-secure subpaths as $\delta$-secure subpaths. In addition, set $$\SecuritySize(\delta):=\SecuritySize(\delta,2t+1).$$
\def\APPENDSECURE{
Fix any identity assignment $\id$ and let $S \subset P$ be a subpath of length $d \geq  \SecuritySize(\delta,\lambda)$, where $S=(v_1, \ldots, v_d)$. In what follows, we may override the nodes $v_i$ with their indices in $S$, namely $i$. Let $r_{\ell}=\lceil (\lambda-1)/2 \rceil$. Given a vertex $v_i \in S$, for $i \in (r_{\ell},d-{r_{\ell}})$, let $B_i$ be the subpath of its $r_{\ell}$ neighborhood in $S$. Formally, $B_i=[i-r_{\ell},i+r_{\ell}]$.  In addition, let $\mathcal{B}_i$ be the event that all nodes in $B_i$ say \yes~ when applying $\distalg$ on $P$. That is $\mathcal{B}_i=\cE(P,\inp,\id,V(B_i))$.
Let $$R=(r_{\ell}+t+1,d-{r_{\ell}}-t)$$ be the ranges of indices $i$ whose $r_{\ell}$ neighborhood $B_i$ is internal in $S$, i.e., $B_i \subset [t+1, d-t-1]$. That is by definition, for every $i \in R$, it holds that $B_i$ is at least of length $\lambda$ and is internal in $S$. See Fig. \ref{fig:fact} for illustration.
We would like to show that there exists $i^{*} \in R$, such that $B_{i^{*}}$ is a $(\delta,\lambda)$-secure subpath, i.e., that $\Pr[\mathcal{B}_i]\geq 1-\delta$.
\par To establish the existence of $i^{*} \in R$, we bound from above the size of the set $\mathcal{H}$ containing all indices $j \in R$ whose $B_j$ is \emph{not} a $(\delta,\lambda)$-secure subpath.
Formally, define
$$\mathcal{H}=\{j \in R \mid \Pr[\mathcal{B}_j] < 1-\delta\}.$$
We next upper bound the size of $\mathcal{H}$, by covering the integers in $R$ by at most $$Q=2(t+r_{\ell})+1$$ sets, each being a $Q$-independent set. That is, every two integers in the same set, are at least $Q$ apart in $P$. Let $\mathcal{J}_1, \ldots, \mathcal{J}_{Q}$ be the $Q$-independent sets that cover the indices of $R$.
\par Specifically, for $s \in [1, Q]$ and $m(S)=\lceil (d-4(r_{\ell}+t))/Q\rceil$, we define
$$\mathcal{J}_{s}=\{s+r_{\ell}+t+1+j \cdot Q \mid  j\in [0,m(S)]\} \cap [1,n].$$ Observe, that as desired, $R \subset \bigcup_{s \in [1, Q]} \mathcal{J}_{s}$, and for each $s \in [1, Q]$, $\mathcal{J}_{s}$ is a $Q$-independent set. In what follows, fix $s \in [1, Q]$ and let $\mathcal{J}=\mathcal{J}_{s}$.
Let $\mathcal{J}'=\mathcal{J} \cap \mathcal{H}$, the indices in $\mathcal{J}$ whose $r_{\ell}$ neighborhood does not correspond to  $(\delta,\lambda)$-secure subpath.
On the one hand, since to a $(P,\inp) \in \cL$, we have that
\begin{equation}
\label{eq:sec_up}
\Pr \left[\bigcup_{i \in \mathcal{J}'} \mathcal{B}_i \right]\geq p.
\end{equation}
On the other hand, observe that for every two indices $i_1, i_2 \in \mathcal{J}'$, it holds that the distance between every $u \in B_{i_1}$ and $v \in B_{i_2}$ is at least $2t+1$ (since  $J'$ is $Q$-independent). Hence, the events $\mathcal{B}_{i_1}$ and $\mathcal{B}_{i_2}$ are independent. We therefore get that
\begin{equation}
\label{eq:sec_down}
\Pr \left[\bigcup_{i \in \mathcal{J}'} \mathcal{B}_i \right]=\prod_{i \in \mathcal{J}'} \Pr \left[\mathcal{B}_i \right] < (1-\delta)^{|\mathcal{J}'|},
\end{equation}
where the last inequality follows by the fact that $\mathcal{J}' \subset \mathcal{H}$.
Combining Eq. (\ref{eq:sec_up}) and (\ref{eq:sec_down}) we get that $$|\mathcal{J}'| < \log p/ \log(1-\delta).$$ We are now ready to upper bound the size of $\mathcal{H}$.
Noting that $S$ can be covered by an union of $Q$ sets $\mathcal{J}_s$ each of which is $Q$-independent, we get that
\begin{equation}
\label{eq:sec_bad}
|\mathcal{H}|=\sum_{i=1}^{Q} |\mathcal{J}'_i|< Q \cdot \frac{\log p}{\log(1-\delta)},
\end{equation}
where $\mathcal{J}'_i=\mathcal{J}_i \cap \mathcal{H}$.
Finally note that $|R| > |\mathcal{H}|$. This follows by the fact that
$|R|=|S|-2(r_{\ell}+t+1)$ and $|S|\geq \SecuritySize(\delta,\lambda)$. Hence, \begin{eqnarray*}
|R|&=&|S|-2(r_{\ell}+t+1)
\\&\geq&
\SecuritySize(\delta,\lambda)-2(r_{\ell}+t+1)
\\&\geq&
Q \cdot \frac{\log p}{\log(1-\delta)}
\\&>&
|\mathcal{H}|,
\end{eqnarray*}
where the last inequality follows by Eq. (\ref{eq:sec_bad}).
By the pigeonhole principle we get that there exists $i^{*} \in R$ which is not in $\mathcal{H}$. Thus $B_{i^{*}}$ is a $(\delta, \lambda)$-secure subpath, as required. The fact follows.
}
\def\APPENDFIGFACT{
\begin{figure}[hb]
\begin{center}
\includegraphics[scale=0.3]{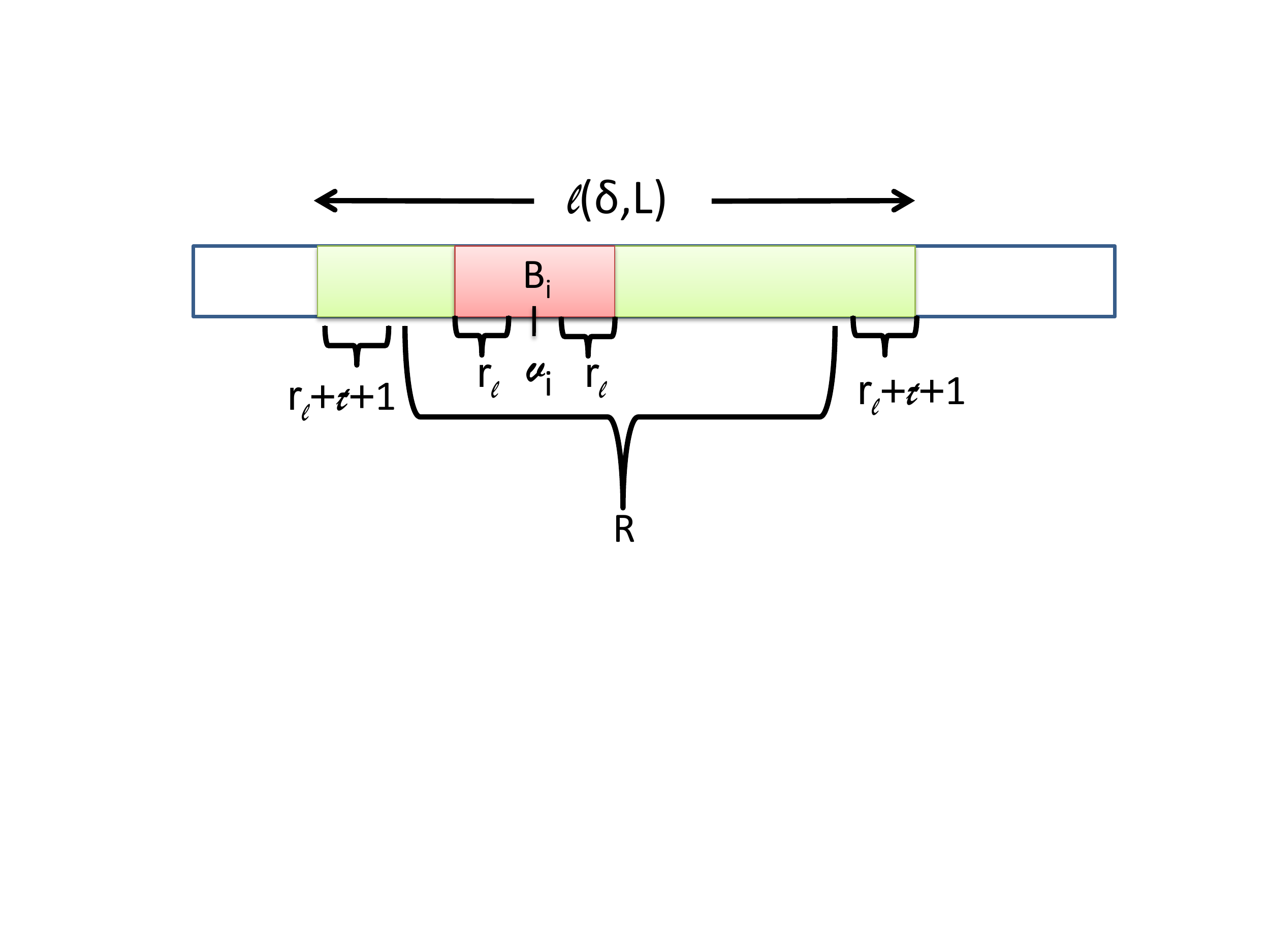}
\caption{ \label{fig:fact}
Presented is path $P$. The green $\SecuritySize(\delta,\lambda)$-length segment corresponds to $S$. The red subpath $B_i$ whose center is $v_i$ is a $(\delta,\lambda)$-secure subpath candidate, as $i \in R$.
}
\end{center}
\end{figure}
}
Let us next illustrate a typical use of Fact~\ref{fact-secure}. Recall that $t$ denotes the running time of the $(p,q)$-decider $\distalg$ on $(P,\inp) \in \cL$ with IDs $\id$. Let $S$ be a subpath of $P$ of length $\SecuritySize(\delta)$. Denote by $L$ (resp., $R$) the subpath of $P$ to the ``left'' (resp., ``right'') of $S$. Informally, if the length of $S$ is larger than $2t+1$, then $S$ serves as a separator between the two subpaths $L$ and $R$. This follows since as algorithm $\distalg$ runs in $t$ rounds, each node in $P$ is affected  only by its $t$ neighborhood. As the $t$ neighborhood of every node $u \in L$ and $v \in R$ do not intersect, the events
$\cE(P,\inp,\id,L)$ and  $\cE(P,\inp,\id,R)$ are independent.

The secureness property becomes useful when bounding the probability that at least some node in $P$ says \no.  A natural approach to upper bound this probability is by applying a union bound on the events $\cE \left(P,\inp,\id,V(L) \cup V(R) \right)$ and $\cE \left(P,\inp,\id,V \left(S \right) \right)$.
Letting $\cE'$ denote the event complementary to $\cE(P,\inp,\id,V(P))$,
we have
\begin{eqnarray*}
\Pr[\cE'] &=& 1-\Pr[\cE(P,\inp,\id,V(P))]
\\ &\leq&
(1-\Pr[\cE(P,\inp,\id,V(L))]
\\ &&
\hbox{\hskip 20pt} \cdot \Pr[\cE(P,\inp,\id,V(R))])
\\ &&
+~ (1-\Pr[\cE(P,\inp,\id,V(S))])
\\ & \leq &
1-\Pr[\cE(P,\inp,\id,V(L))]
\\&&
\hbox{\hskip 20pt} \cdot \Pr[\cE(P,\inp,\id,V(R))]+\delta~.
\end{eqnarray*}
The specific choice of $\lambda$ and $\delta$ depends on the context. Informally, the guiding principle is to set $\delta$ small enough so that the role of the central section $S$ can be neglected, while dealing separately  with the two extreme sections $L$ and $R$ become manageable for they are sufficiently far apart.

\section{The $B_k$ hierarchy is strict}

In this section we show that the classes $B_k$, $k\geq 1$, form an infinite hierarchy of distinct classes, thereby proving that the general ability to boost the probability of success for a randomized decision problem is quite limited. In fact, we show separation in a very strong sense: there are decision problems in $B_{k+1}(0)$, i.e., that have a $(p,q)$-decider running in zero rounds with $p^{1+1/(k+1)}+q>1$, which cannot be decided by a $(p,q)$-decider with $p^{1+1/k}+q>1$, even if the number of rounds of the latter is as large as $n^{1-\varepsilon}$ for every fixed $\varepsilon>0$.

\begin{theorem}
\label{thm:seperation}
$B_{k+1}(0)  \setminus B_{k}(t) \neq \emptyset $ for every $k \geq 1$ and every $t=o(n)$.
\end{theorem}

\Proof
Let $k$ be any positive integer. We consider the following distributed language, which is a generalized variant $\leaderk$ of the problem $\leader$ introduced in~\cite{FKP11}. As in $\leader$, the input $\inp$ of $\leaderk$ satisfies $\inp\in\{0,1\}^n$, i.e., each node $v$ is given as input a boolean $\inp(v)$.  The language \leader-$k$ is then defined by:
\vspace*{-1ex}
\begin{eqnarray*}
\mbox{\tt At-Most-k-Selected } (\mbox{\leaderk}) &=&
\\
\{(G,\inp)  \;\mbox{s.t.}\; \parallel\inp\parallel_{_1}\;\leq k\}.
\end{eqnarray*}
Namely, $\leaderk$ consists of all instances containing at most $k$ selected nodes (i.e., at most $k$ nodes with input~1), with all other nodes unselected (having input~0). In order to prove Theorem~\ref{thm:seperation}, we show that $\leaderk \in B_{k+1}(0) \setminus B_{k}(t)$ for every $t=o(n)$.

We first establish that $\leaderk$ belongs to $B_{k+1}(0)$. We adapt  algorithm $\distalg$ presented in~\cite{FKP11} for $\leader$ to the case of $\leaderk$. The following simple randomized algorithm runs in $0$ time: every node $v$ which is not selected, i.e., such that $\inp(v)=0$,   says \yes���½\/ with probability $1$; and every node which is selected, i.e., such that $\inp(v)=1$, says ���½\yes\/ with probability $p^{1/k}$, and \no\/ with probability $1-p^{1/k}$. If the graph has $s\leq k$ nodes selected, then all nodes say \yes\/  with probability $p^{s/k}\geq p$, as desired. On the other hand, if there are $s \geq k+1$ selected nodes, then at least one node says \no\/ with probability $1-p^{s/k} \geq 1-p^{(k+1)/k}=1-p^{1+1/k}$. We therefore get a $(p,q)$-decider with $p^{1+1/k}+q\geq 1$, that is, such that $p^{1+1/(k+1)}+q > 1$. Thus $\leaderk\in B_{k+1}(0)$.

We now consider the harder direction, and prove that $\leaderk \notin B_{k}(t)$, for any $t=o(n)$. To prove this separation, it is sufficient to consider $\leaderk$ restricted to the family of $n$-node paths. Fix a function $t=o(n)$, and assume, towards contradiction, that there exists a distributed $(p, q)$-decider $\distalg$ for $\leaderk$ that runs in $O(t)$ rounds, with $p^{1+1/k}+q>1$.  Let $\varepsilon \in (0,1)$ be such that $p^{1+1/k+\varepsilon}+q>1$. Let $P$ be an $n$-node path, and let $S\subset P$ be a subpath of $P$.
Let $\delta \in [0,1]$ be a constant satisfying
\begin{equation}
\label{eq:lambda}
0 <\delta < p^{1+1/k}\left( 1-p^{\varepsilon} \right) /k~.
\end{equation}
Consider a positive  instance and a negative instance of $\leaderk$, respectively denoted by
\[
I =(P, \inp) \; \mbox{and} \; I'=(P, \inp')~.
\]
Both instances are defined on the same $n$-node path $P$, where
$
n \geq k\left(\SecuritySize(\delta)+1\right)+1.
$

Recall that $\SecuritySize(\delta)=\SecuritySize(\delta,2t+1)$ (see Eq. (\ref{eq:defsecuritysize})).
We consider executions of $\distalg$ on these two instances,
where nodes are given the  same id's.
Both instances have almost the same input. In particular,
the only difference is that instance $I$ contains $k$ selected nodes,
whereas $I'$ has the same selected nodes as $I$ plus one additional selected node.
Therefore $I$ is legal, while $I'$ is illegal.
In $I'$, the path $P$ is composed of $k+1$ sections, each containing a unique selected node,
and where each pair of consecutive sections separated by a $\delta$-secure subpath.
More precisely, let us enumerate the nodes of $P$ from $1$ to $n$, with node $v$ adjacent to nodes $v-1$ and $v+1$, for every $1<v<n$. Consider the $k$ subpaths of $P$ defined by:
\[S_i=[(i-1)\SecuritySize(\delta)+i+1,i \cdot \SecuritySize(\delta)+i]\]
for $i=\{1, \ldots, k\}$. Let the selected nodes in $I'$ be positioned as follows. Let $u_1=1$ and let $u_i=(i-1)\SecuritySize(\delta)+i$ for $i=2, \ldots, k+1$. Then set
\[
\inp'(v)=\left\{\begin{array}{ll}
1, & \mbox{if $v=u_i$ for some $i \in \{1, ..., k+1\}$}\\
0, & \mbox{otherwise.}
\end{array}\right.
\]
See Fig. \ref{fig:amos_k}(a) for a schematic representation of $I'$.
Our next goal is to define the legal instance $I=(P,\inp)$. To do so, we begin by claiming that each $S_i$ contains a $\delta$-secure internal subpath $S_i'=[a_i,b_i]$. Naturally, we would like to employ Fact~\ref{cl:sec_zone_guarantee}. However, Fact~\ref{cl:sec_zone_guarantee} refers to subpaths of \emph{valid} instances $(P,\inp) \in \cL$, and $I'$ is illegal. So instead, let us focus on the instance $(S_i, \inp'_{S_i})$. Since $(S_i, \inp'_{S_i})$ contains no leaders, $\parallel\inp'_{S_i}\parallel_{_1}\;= 0$, it follows that $(S_i, \inp'_{S_i}) \in \cL$, and Fact~\ref{cl:sec_zone_guarantee} can be applied on it. Subsequently, since $|S_i|> \SecuritySize(\delta)$ it follows that $S_i$ contains an internal $\delta$-secure subpath $S'_{i}=[a_i,b_i]$, whose $t$ neighborhood is strictly in $S_i$.
Therefore, when applying algorithm $\distalg$ on $(S_i, \inp'_{S_i},\id_{S_i})$ and on $(P, \inp',\id)$, the nodes in the $(2t+1)$-length segment $S'_i$ behave the same, thus $\Pr[\cE(P,\inp',\id,V(S'_i))]=\Pr[\cE(S_i,\inp'_{S_i},\id_{S_i},V(S'_i))]$. Hence, $S'_i$ is a $\delta$-secure subpath in $(P,\inp',\id)$ as well, for every $i \in \{1, ..., k\}$, see Fig. \ref{fig:amos_k}(b).
\par The $\delta$-secure subpaths $S'_i$'s are now used to divide $P$ into $2k+1$ segments. Specifically, there are $k+1$ segments $T_{i}$, $i=1,\dots,k+1$, each with one selected node. The $\delta$-secure subpaths $S'_i=[a_i,b_i]$ separate $T_{i}$ from $T_{i+1}$. More precisely, set $T_{1}=[1, a_1-1]$, $T_i=[b_{i-1}+1, a_i-1]$ for $i \in {2, ..., k}$, and
$T_{k+1}=[b_{k}+1,n]$, getting \[P= T_{1} \circ S'_1 \circ T_2 \circ S'_2 \circ \ldots \circ T_k \circ S'_k \circ T_{k+1}\]
where $\circ$ denotes path concatenation.
Let $\mathcal{T}_i=\cE(P,\inp',\id,V(T_i))$ be the event that all nodes in
the subpath $T_{i}$ say \yes\/ in the instance $I'$, for $i\in\{1, ..., k+1\}$ and let $p_i=\Pr[\mathcal{T}_i]$ be
its probability. Let $j$ be such that $p_{j}=\max_i p_i$.
We are now ready to define the valid instance $I=(P, \inp)$:
\[
\inp(v) = \left\{\begin{array}{ll}
1, &  \mbox{if $v=u_i$ for some $i \in \{1, ..., k+1\}$,} \\
  &  i\neq j \\
0, &   \mbox{otherwise.}
\end{array}\right.
\]
Note that $\parallel\inp'\parallel_{_1}\;= k+1$ and $\parallel\inp \parallel_{_1}\;= k$, thus $I \in \leaderk$ while $I'\notin\leaderk$.
See Fig. \ref{fig:amos_k}(c,d) for an illustration of $I$ versus $I'$. 
\par We now make the following observation.
\begin{claim}
$\forall i \neq j, ~\Pr[\cE(P,\inp,\id,V(T_i))]=p_i$.
\end{claim}
\Proof
This follows since the distance between any two nodes $u$ (resp., $v$) in distinct  $T_{i}'s$ is greater than $t$, which implies that $\inp(L_i \circ T_i \circ R_i) = \inp'(L_i \circ T_i \circ R_i)$ where $L_i$ (resp., $R_i$) is the subpath of length $t$ to the left (resp., to the right) of $T_i$ in $P$, from which it follows that under $\distalg$ the nodes of $T_i$ have the same behavior in both instances $I$ and $I'$.
\QED
Let $\cN$ (resp., $\cN'$) be the event that there exists at least one node in $I$ (resp., $I'$) that says \no\/ when applying algorithm $\distalg$. Similarly, let $\cY$ (resp., $\cY'$) be the event stating that all nodes in the configuration $I$ (resp., $I'$) say \yes\/. Let $\mathcal{T}=\bigcup_{i=1}^{k+1}\mathcal{T}_i$ be the event that all nodes in each subpaths $T_{i}$, for $i \in \{1,...,k+1\}$ say \yes\/ in the instance $I'$. For every $i \in \{1, ..., k\}$, let $\mathcal{S}_i=\cE(P,\inp',\id,V(S'_i))$ be the event that all nodes in the $\delta$-secure subpath $S'_i$ say \yes\/ in the instance $I'$.
We have $\Pr(\cY)=\Pr[\cE(P,\inp,\id,V(P))] \; \mbox{and} \; \Pr(\cY')=\Pr[\cE(P,\inp',\id,V(P))$,
while $\Pr(\cN)=1-\Pr(\cY)$ and $\Pr(\cN')=1-\Pr(\cY')$.
\par Since $\distalg$ is a $(p,q)$-decider, as we assume by contradiction that $\leaderk$ in $B_k$, we have $\Pr(\cN') \geq q$, and thus $\Pr(\cN') > 1-p^{1+1/k+\varepsilon}$. Therefore, $\Pr(\cY') < p^{1+1/k+\varepsilon}$. Moreover,  since $I \in \leaderk$, we also have that $\Pr(\cY)\geq p$. Therefore, the ratio ${\hat\rho} = \Pr(\cY') / \Pr(\cY)$ satisfies
\begin{equation}
\label{eqn:ratio_k_upper}
{\hat\rho}
< p^{1/k+\varepsilon}~.
\end{equation}
On the other hand, note that by applying the union bound to the $k+1$ events $\mathcal{T},\bigcup_{i=1}^{k}\mathcal{S}_{i}$, we get
\begin{eqnarray*}
\Pr(\cN') &\leq& \left(1-\Pr[\mathcal{T}] \right)+\left(\sum_{i=1}^{k}(1-\Pr[\mathcal{S}_{i}]) \right)
\\ &\leq&
1-p_j \cdot \prod_{i \neq j} p_i +k \cdot \delta,
\end{eqnarray*}
where the last inequality follows by the fact that each $S'_i$ is a $(\delta,2t+1)$-secure subpath, thus the events $\mathcal{T}_{i_1},\mathcal{T}_{i_2}$ are independent for every $i_1, i_2 \in \{1, ..., k+1\}$ (since the distance between any two nodes $u \in T_{i_1}$ and $v \in T_{i_2}$ is at least  $2t+1$).  This implies that
$
\Pr(\cY') \geq p_j \cdot \prod_{i \neq j} p_i -k \cdot \delta~.
$
Since $\Pr(\cY) \leq \prod_{i \neq j} p_i$ (by the independence of the events $\mathcal{T}_{i_1},\mathcal{T}_{i_2}$, for every $i_1, i_2 \in \{1,...k+1\}$),
it then follows that the ratio $\hat\rho$
satisfies
\begin{eqnarray}
{\hat\rho} &\geq&
\frac{p_j \cdot \prod_{i \neq j} p_i -k \cdot \delta}{\prod_{i \neq j} p_i} \nonumber
\\&\geq&
p_j -\frac{k \cdot \delta}{\prod_{i \neq j} p_i}
\geq
p_j -k \cdot\delta/p~,
\label{eqn:ratio_k_lower}
\end{eqnarray}
where the last inequality follows by the fact that $I \in \leaderk$ and
thus $\prod_{i \neq j} p_i \geq \Pr(\cY) \geq p$.
Finally, note that $p_j \geq p^{1/k}$. This follows since
$p_j \geq p_{i}$ for every $i \in \{1, ..., k+1\}$, so $p_j^k \geq \prod_{i \neq j} p_i \geq p$. By Eq.~(\ref{eqn:ratio_k_lower}), we then have that ${\hat\rho} \geq p^{1/k} -k \cdot\delta/p$. Combining this with Eq.~(\ref{eqn:ratio_k_upper}), we get that
$
p^{1/k} -k \cdot \delta/p < p^{1/k+\varepsilon}~,
$
which is in contradiction to the definition of $\delta$ in Eq. (\ref{eq:lambda}).
\QED
\def\APPENDFIGAMOS{
\begin{figure}[hb]
\begin{center}
\includegraphics[scale=0.3]{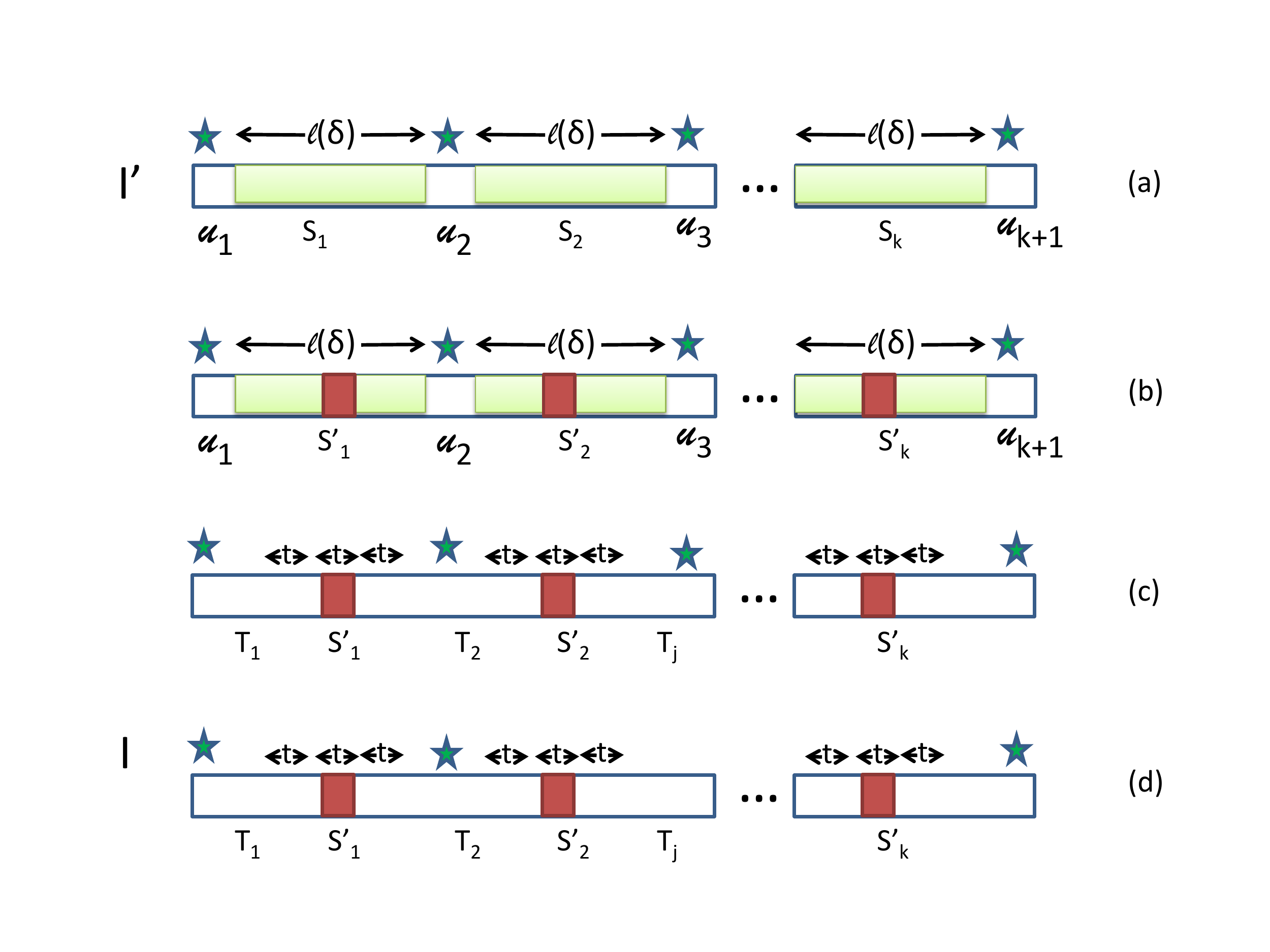}
\caption{ \label{fig:amos_k} Illustration of the constructions for Theorem~\ref{thm:seperation}.
\sf
(a) The instance $I'=(P,\inp')$ with $k+1$ leaders separated by $\SecuritySize(\delta)$-length segements, $S_i$.
(b) The $\delta$-secure subpaths $S'_i$ in each $S_i$ are internal to $S_i$.
(c) The leader-segments $T_i$ interleaving with $\delta$-secure subpaths $S'_i$.
(d) The legal instance $I=(P,\inp)$, the $j^{th}$ leader of $I'$ is discarded, resulting in a $k$ leader instance.
}
\end{center}
\end{figure}
}


Finally, we show that the $B_{k}(t)$ hierarchy does not capture all languages even for $k=\infty$ and $t$ as large as $o(n)$. The proof of the following theorem is deferred to the Appendix.
\begin{theorem}
\label{thm:seperationinfty}
There is a language not in $B_\infty(t)$, for every $t=o(n)$.
\end{theorem}
\def\APPENDINFINITY{
We exhibit one specific language not in $B_\infty(t)$, for every $t=o(n)$. This language consists of determining whether the underlying network is acyclic. Specifically, let $$\tree=\{(G,\epsilon) \mid G\;\mbox{is a tree}\},$$
where $\epsilon$ is the null input. Fix a function $t=o(n)$. Assume, towards contradiction, that there exists some finite $k$ such that  $\tree\in B_k(t)$. Then there is a $(p,q)$-decider for $\tree$, given by $\distalg$, running in $t$ rounds, with $p^{1+1/k}+q-1>0$. Hence, in particular, there  exists some $\varepsilon >0$ such that $p+q-1 >\varepsilon$. Define
\begin{equation}
\label{eq:delta_tree}
\delta=\varepsilon < p+q-1.
\end{equation}
We consider graphs $G$ of size $n>\left \lceil \frac{21 \cdot \log p} {\log(1-\delta)} \right \rceil$. We will show that $\tree \notin B_{k}(t)$ for any
\begin{equation}
\label{eq:t_tree}
t \leq \left \lfloor \frac{\log(1-\delta)}{21 \cdot \log p} \right \rfloor \cdot n=o(n).
\end{equation}
 Consider the cycle $C$ with $n$ nodes labeled consecutively from $1$ to $n$, and the path $P$ with nodes labeled consecutively from $1$ to $n$. This labeling defines the identity assignment $\id^1$. In the input configuration $(P,\epsilon)$, the probability that  all nodes say \yes\/ when executing $\distalg$ is at least $p$. Let
us identify a subpath $S=[x-t, \ldots, x+t+1]$ of $P$ to be used as an internal $(\delta,2(t+1))$-secure subpath in $P$. I.e.,
\begin{equation}
\label{eq:prob_s}
\Pr[\cE(P,\epsilon,\id^1,V(S))] \geq 1- \delta~.
\end{equation}
Note that, by Eq.~(\ref{eq:defsecuritysize}, \ref{eq:delta_tree}, \ref{eq:t_tree}), it follows that $n>\SecuritySize(\delta,2(t+1))$. Hence by Fact~\ref{cl:sec_zone_guarantee}, since $(P,\epsilon)\in \cL$, there exists such internal  subpath $S \subset P$. Consider the event $\cE(P,\inp,\id,V(S))$ stating that  all nodes in subpath $S$ of $P$ with input $\inp$ and identity-assignment $\id$ return \yes.
We have that
\begin{equation}
\label{eq:claimQ1}
\Pr[\cE(P,\epsilon,\id^1,V(S))]=\Pr[\cE(C,\epsilon,\id^1,V(S))]~.
\end{equation}
Consider a subpath $\widehat{S}$ composed of the subpath $S$ padded with a block $L$ of $t$ nodes before it and a block $R$ of $t$ nodes after it. Indeed, since $S$ is an internal subpath of $P$ (i.e., it is at distance at least $t+1$ from $P$'s endpoints), the set of nodes of $\widehat{S}=L \circ S \circ R=[x-2t, \ldots, x+2t+1]$ appears consecutively in both $P$ and in $C$ with identity assignment $\id^1$, and  $\widehat{S}$ have the same identities (with $\id^1$) and degrees in both $C$ and $P$.
We now consider another identity-assignment $\id^2$ for $P$, with nodes labeled consecutively from $x+1$ to $n$, and then from $1$ to $x$. Consider the $(n-2(t+1))$-node subpath $$S'=[x+t+2, \ldots, n, 1, \ldots, x-t-1]~.$$ We have
\begin{equation}
\label{eq:claimQ2}
\Pr[\cE(P,\epsilon,\id^2,V(S'))]=\Pr[\cE(C,\epsilon,\id^1,V(S'))]~,
\end{equation}
Consider a subpath $\widehat{S'}$ composed of the subpath $S'$ padded with a block $L'$ of $t$ nodes before it and a block $R'$ of $t$ nodes after it, i.e., $\widehat{S'}=L' \circ S' \circ R'$. Indeed,  the set of $\widehat{S'}$ nodes appears consecutively in both $C$ and $P$ with identity assignment $\id^2$, and  $L' \circ S' \circ R'$ have the same identities (with $\id^2$) and degrees in both $C$ and $P$, where $L'$ (resp., $R'$) is the subpath composed of  the $t$ nodes with identities $\id^2$ immediately larger than $x+t+1$ (resp., smaller than $x-t$). Formally, we have that $\id^1_{\widehat{S'}}=\id^2_{\widehat{S'}}$.
See Fig. \ref{fig:tree} for illustration.
Let $\mathcal{S}=\cE(C,\epsilon,\id^1,V(S))$, (resp., $\mathcal{S'}=\cE(C,\epsilon,\id^1,V(S'))$)  be the event
that all nodes of $S \subset C$ (resp., $S' \subset C$) say \yes.
We can now combine these previous results to derive a contradiction. Since $C \notin \tree$,  by applying the union bound on the events $\mathcal{S}$ and $\mathcal{S'}$, and using Eq. (\ref{eq:claimQ1}) and (\ref{eq:claimQ2}), we get that
\begin{eqnarray*}
q &\leq& 1-\Pr[\cE(C,\epsilon,\id^1,V(C))]
\\&\leq&
\left(1-\Pr[\mathcal{S'}]\right)+\left(1-\Pr[\mathcal{S}] \right)\\
& = & \left(1-\Pr[\cE(P,\epsilon,\id^2,V(S'))]\right)
\\&& +
\left(1-\Pr[\cE(P,\epsilon,\id^1,V(S))]\right)
\\&\leq &
\left(1-\Pr[\cE(P,\epsilon,\id^2,V(S'))]\right)+\delta \label{ineq:1}
\end{eqnarray*}
where the last inequality holds by Eq.~(\ref{eq:prob_s}). Therefore we get
\[
q \leq 1-p+\delta~,   \label{ineq:2}
\]
by noticing that  $\Pr[\cE(P,\epsilon,\id^2,V(S'))]\geq p$  since $P \in \tree$. Finally, by Eq. (\ref{eq:delta_tree}), we eventually get $q < 1-p+p+q-1$ or $q < q$, contradiction.
\QED
}

\def\APPENDFIGTREE{
\begin{figure}
\begin{center}
\includegraphics[scale=0.5]{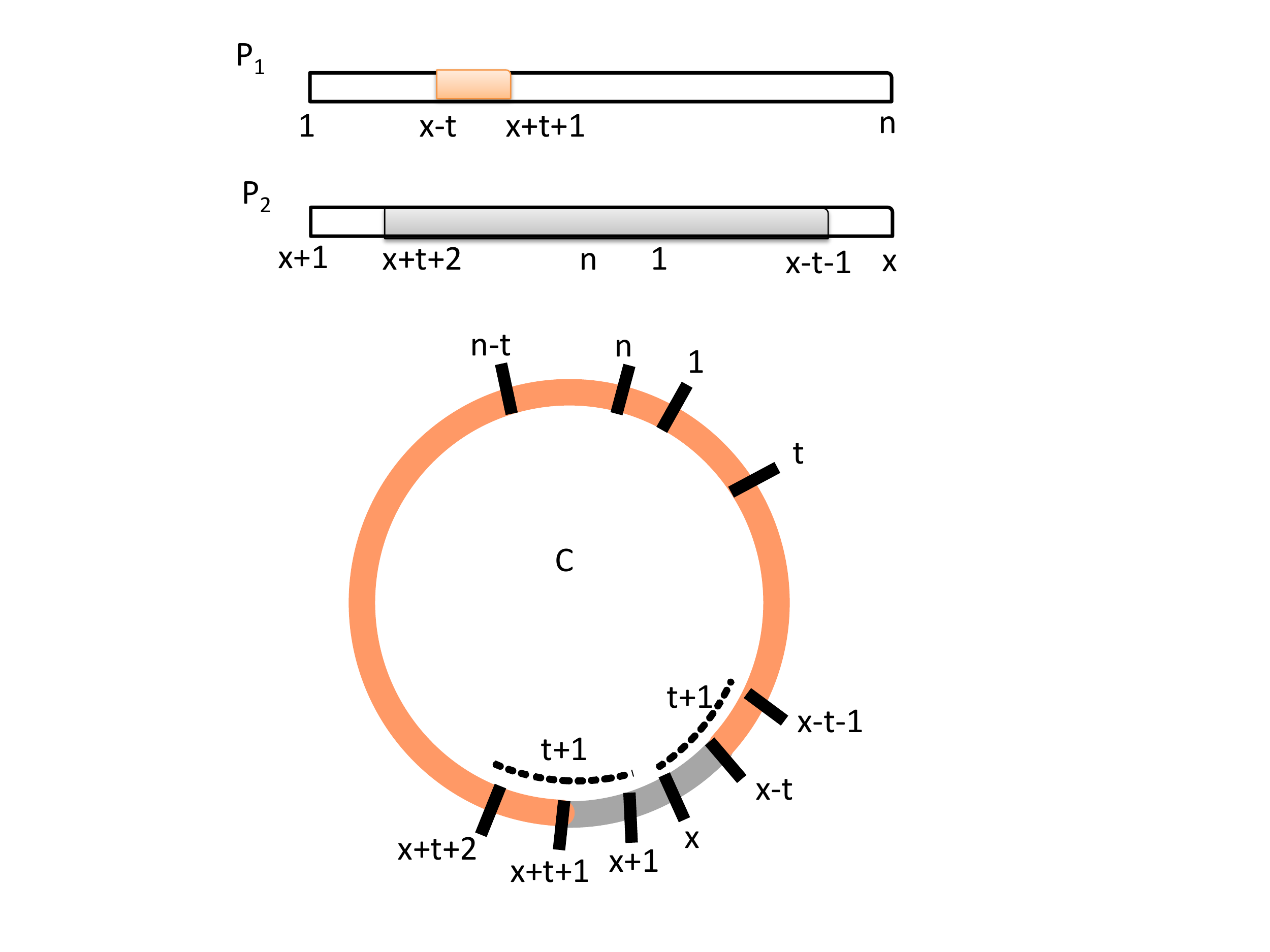}
\caption{\label{fig:tree} Illustration of the constructions for Theorem~\ref{thm:seperationinfty}.
Shown are paths $P_1$,$P_2$ and cycle $C$.
When applying Algorithm $\distalg$ on path $P_1$ (respectively, $P_2$) and on cycle $C$, the nodes in the segment $[x-t, x+t+1]$ (resp., $[x+t+2, \ldots,1,\ldots, x+t+1]$) behave the same.
}
\end{center}
\end{figure}
}
\section{A sharp threshold between determinism and randomization}

It is known~\cite{FKP11} that beyond the threshold $p^2+q=1$, randomization does not help. This result however holds only for a particular type of languages, called  \emph{hereditary}, i.e., closed under inclusion. In this section, we provide one more evidence supporting our belief that the threshold $p^2+q=1$ identified in~\cite{FKP11} holds for \emph{all} languages, and not only for hereditary languages. Indeed, we prove that, restricted to path topologies and finite inputs, \emph{every} language $\cL$ for which there exists a $(p,q)$-decider running in constant time, with $p^2+q>1$,
can actually be decided deterministically in constant time.

\begin{theorem}
\label{thm:hard-direction}
Let $\cL$ be a distributed language restricted to paths, with a finite set of input values. If $\cL \in B_1(O(1))$, then $\cL\in \LD(O(1))$.
\end{theorem}

\Proof
Let $\cL\in B_1(O(1))$ be a distributed language restricted to paths, and defined on the (finite) input set $\Alphabet$.
Consider a distributed $(p, q)$-decider $\distalg$ for $\cL$ that runs in $t=O(1)$ rounds, with $p^{2}+q>1$.
Fix a constant $\delta$ such that $0<\delta<p^2+q-1$.
\par Given  a subpath $S$ of a path $P$, let us denote by $S_{l}$ (respectively, $S_{r}$)
the subpath of $P$ to the left (resp., right) of $S$, so that
$P = S_{l}\circ S \circ S_{r}$.

Informally, a collection of three paths $P, P'$, and $P''$ (of possibly different lengths) is called
a \emph{$\lambda$-path triplet} if (1) the inputs of those paths agree on some ``middle'' subpath of size at least $\lambda$, (2) paths $P$ and $P''$ coincide on their corresponding ``left'' parts, and (3) paths
$P'$ and $P''$ coincide on their ``right'' parts. See~Figure~\ref{path-triplet}.
Formally, a {\em $\lambda$-path triplet}
is a triplet
$[(P,S,\inp), (P',S', \inp'), (P'',S'', \inp'')]$ such that
$|P|, |P'|,|P''| \geq \lambda$,
$\inp,\inp',\inp''$ are inputs on these paths, respectively,
and $S\subset P$, $S'\subset P'$, $S''\subset P''$ are three subpaths satisfying (1) $|S|=|S'|=|S''|\geq \lambda$,
(2) $\inp_S=\inp'_{S'}=\inp''_{S''}$, and (3)
$\inp''_{S''_{l}}=\inp_{S_{l}}$ and $\inp''_{S''_{r}}=\inp'_{S'_{r}}$.
The proof of the following claim is deferred to the Appendix.
\def\APPENDFIGPATHS{
\begin{figure}
\begin{center}
\includegraphics[scale=0.5]{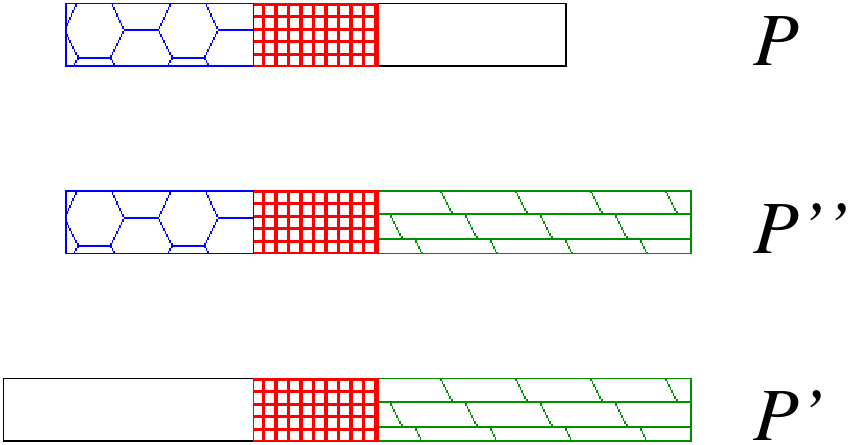}
\caption{Example of a $\lambda$-path triplet (the red zone is of length at least $\lambda$).}
\label{path-triplet}
\end{center}
\end{figure}
}

\begin{claim}\label{claim-path}
Let $[(P,S,\inp), (P',S', \inp'), (P'',S'', \inp'')]$ be a $\lambda$-path triplet. If $\lambda \geq \SecuritySize(\delta)$, for ~$\ell$ as defined in Eq.~(\ref{eq:defsecuritysize}), then
$
\Big ((P,\inp)\in \cL  \; \mbox{\rm and} \; (P',\inp')\in \cL\Big ) \; \Rightarrow \; (P'',\inp'')\in \cL.
$
\end{claim}
\def\APPENDPATHTRIPLE{
Consider an identity assignment $\id''$ for $(P'',\inp'')$. Let  $\id$ and $\id'$ be
identity assignments  for  $(P,\inp)$, and $(P',\inp')$, respectively, which agree with $\id''$  on the corresponding nodes.
That is: (a) assignments $\id$, $\id'$, and $\id''$ agree on the nodes in $S$, $S'$ and $S''$, respectively; (b) $\id$ and $\id''$ agree on the nodes in $S_{\ell}$ and $S''_{\ell}$, respectively; and  (c) $\id$ and $\id''$ agree on the nodes in $S'_{r}$ and $S''_{r}$, respectively.
Since $(P,\inp)\in \cL$, and since $|S|=\lambda\geq \SecuritySize(\delta)$,
it follows from Fact~\ref{fact-secure} that $S$ contains an internal $\delta$-secure subpath $H$. Then, let $H'$ and $H''$ be the subpaths of $P'$ and $P''$ corresponding to $H$.
Since $S$ and $S''$ coincide in their inputs and identity assignments,
then $H,H', H''$ have the same $t$-neighborhood in $P,P',P''$ respectively.
Hence, $H''$ is also a $\delta$-secure (when running algorithm $\distalg$ in instance $(P'',\inp'')$).
Since both $(P,\inp)$ and $(P',\inp')$ belong to $\cL$,  we have
 \[
 \Pr[\cE(H''_{\ell},\id'',\inp'')=\Pr[\cE(H_{\ell},\id,\inp)])] \geq p
 \]
 and
 \[
 \Pr[\cE(H''_{r},\id'',\inp'')=\Pr[\cE(H'_{r},\id',\inp')])] \geq p.
 \]
 Moreover, as $|H''|\geq 2t+1$, the two events $\cE(H''_{\ell},\id'',\inp'')$ and $\cE(H''_{r},\id'',\inp'')$ are independent. Hence
\[
\Pr[\cE(H''_{\ell}\cup H''_r,\id'',\inp'')]\geq p^2.
\]
In other words, the probability that some node in $H''_{\ell}\cup H''_r$ says ``no'' is at most $1-p^2$.
It follows, by union bound, that the probability that some node in $H''$ says ``no'' is at most
$1-p^2+\delta< q$. Since $\distalg$ is a $(p, q)$-decider  for $\cL$, it cannot be the case that
$(H'',\inp'')\notin \cL$.
\QED
}
%
%

We now observe that, without loss of generality, one can assume that in all instances $(P,\inp)$ of $\cL$,  the two extreme vertices of the path $P$ have a special input symbol~$\NewSymb$. To see why this holds,
let~$\NewSymb$ be a symbol not in $\Alphabet$, and
consider the following language $\cL'$ defined over $\Alphabet\cup  \{\NewSymb\}$. Language $\cL'$ consists of instances $(P,\inp)$ such that (1) the endpoints of $P$ have input~$\NewSymb$, and (2) $(P',\inp')\in\cL$, where $P'$ is the path resulting from removing the endpoints of $P$, and where $\inp'_v=\inp_v$ for every node $v$ of $P'$. Any $(p,q)$ decider algorithm for $\cL$ (resp., $\cL'$), can be trivially transformed into a $(p,q)$ decider algorithm for  $\cL'$ (resp., $\cL$)  with the same success guarantees and running time. Hence, in the remaining of the proof, we assume that in all instances $(P,\inp)\in\cL$,  the two extreme vertices of the path $P$ have  input~$\NewSymb$.

We say that a given instance $(P,\inp)$ is {\em extendable} if there exists an extension of it in $\cL$, i.e., if there exists an instance $(P',\inp')\in\cL$ such that  $P\subseteq P'$ and $\inp'_{P}=\inp$. The proof of the following claim is deferred to the Appendix.

\begin{claim}\label{claim-extendable}
There exists a (centralized) algorithm $\mathcal{X}$  that, given any configuration $(P,\inp)$ with $|P|\leq  2\SecuritySize(\delta)+1$,  decides whether $(P,\inp)$ is extendable. Moreover, algorithm $\mathcal{X}$ uses messages of constant size.
\end{claim}
\def\APPENDPATHEXTEND{
Observe that since the running time $t$ is constant, then  $2\SecuritySize(\delta)+1$ is also constant.
Therefore, there are only finitely many configurations $(\hat{P},\inp)$ with $|\hat{P}| \le 2\SecuritySize(\delta)+1$ (since $\Alphabet$ is finite). Call this set of configurations $\cal{C}$. Each of the configurations in $\cal{C}$ is either extendable or not. Hence, there exists a function $f:\cC \rightarrow \{0,1\}$ such that for every configuration $C\in \cC$, $f(C)=1$ if and only if $C$ is extendable. This function $f$ can be described in a finite manner, and hence gives rise to an algorithm
as required by the claim.
\QED
}

We may assume, hereafter, that such an algorithm $\mathcal{X}$, as promised by Claim~\ref{claim-extendable},  is part of the language specification given to the nodes.
We show that $\cL\in \LD(O(t))$ by proving the existence of a deterministic algorithm $\cal D$ that recognizes $\cL$ in $O(t)$ rounds. Given a path $P$, an input $\inp$ over $P$, and an identity assignment $\id$, algorithm $\cal D$ applied at a node $u$  of $P$ operates as follows. If $\inp_u=\NewSymb$ then $u$ outputs ``yes''  if and only if  $u$ is an endpoint of $P$. Otherwise, i.e., if $\inp_u\neq\{\NewSymb\}$, then $u$
outputs ``yes''
if and only if $(B_u,\inp_{B_u})$ is extendable (using algorithm $\mathcal{X}$), where $B_u=B(u,\SecuritySize(\delta))$ is the ball centered at $u$, and of radius  $\SecuritySize(\delta)$ in $P$.

Algorithm $\cal D$ is a deterministic algorithm that runs in $\SecuritySize(\delta)$ rounds. We claim that Algorithm $\cal D$ recognizes $\cL$. To establish that claim, consider first an instance $(P,\inp)\in \cL$. For every node $u$, $(P,\inp)\in \cL$ is an extension of $(B_u,\inp_{B_u})$. Therefore, every node $u$ outputs ``yes'', as desired. Now consider an instance $(P,\inp)\notin \cL$. Assume, for the purpose of contradiction, that there exists an identity assignment $\id$ such that, when applying $\cal D$ on $(P,\inp,\id)$, every node $u$ outputs ``yes''.
\begin{claim}
In this case, $|P|>2\SecuritySize(\delta)+1$.
\end{claim}
\Proof
Assume by contradiction that $|P|\leq 2\SecuritySize(\delta)+1$, and consider  the middle node $s$ of $P$. Since $s$ outputs ``yes'', it follows that $(P,\inp)$ can be extended to $(P',\inp')$ such that $(P',\inp')\in \cL$. However, since the extremities of $P$ output ``yes'', it means that their input is~$\NewSymb$. Therefore, as $|P'|>|P|$, we get that there is an internal node of $P'$ which  has input $\NewSymb$, contradicting $(P',\inp')\in \cL$.
\QED
Let $S\subseteq P$ be the longest subpath of $P$ such that there exists an extension $(P',\inp')$ of $(S,\inp_S)$, with $(P',\inp')\in\cL$. Since $|P|>2\SecuritySize(\delta)+1$, and since the middle node of $P$ outputs ``yes'', we have $|S| \geq 2\SecuritySize(\delta)+1$. The proof carries on by distinguishing two cases for the length of $S$.

If $S=P$, then $(P,\inp)$ can be extended to $(P',\inp')\in \cL$. By the same arguments as above, since each extremity $w$ of $P$ has input $\NewSymb$, we conclude that $P=P'$, with $\inp=\inp'$. Contradicting the fact that $(P,\inp)\notin \cL$. Therefore $2\SecuritySize(\delta)+1 \leq |S| < |P|$. Let $a$ and $b$ be such that $S=[a,b]$. As $S$ is shorter than $P$, it is impossible for both $a$ and $b$ to be endpoints of $P$. Without loss of generality, assume that $a$ is not an endpoint of $P$. Since $a$ outputs ``yes'', there exists an extension $(P'',\inp'')\in\cL$ of $(B_a,\inp_{B_a})$. In fact, $(P'',\inp'')$ is also an extension of $\inp_{[a, a+\SecuritySize(\delta)]}$. Since $\inp'$ and $\inp''$ agree on $[a, a+\SecuritySize(\delta) ]$, and since both $(P',\inp')$, and $(P'',\inp'')$ are in $\cL$, we get from Lemma~\ref{claim-path} that $\inp_{[a-1,b]}$ can be extended to an input $(P''',\inp''') \in \cL$, which  contradicts the choice of $S$. The theorem follows.
\QED

\section{On the impossibility of boosting}
Theorems~\ref{thm:seperation} and~\ref{thm:seperationinfty} demonstrate that boosting the probability of success might be doable, but only from $(p,q)$ satisfying $p^{1+1/(k+1)}+q>1$ to $(p,q)$ satisfying $p^{1+1/k}+q>1$ (with the extremes excluded). In this section, we prove that once the inputs may be restricted in certain ways, the ability to boost the success probability become almost null. More precisely, recall that so far we considered languages as collections of pairs $(G,\inp)$ where $G$ is a (connected) $n$-node graph and $\inp\in\Sigma^n$ is the input vector to the nodes of $G$, in some finite of infinite alphabet~$\Sigma$, that is, $\inp(v)\in\Sigma$ for all $v\in V(G)$. An instance of an algorithm $\distalg$ deciding a language $\cL$ was defined as \emph{any} such pair $(G,\inp)$. We now consider the case where the set of instances is restricted to some specific subset of inputs $\cI \subset \Sigma^n$. That is, the distributed algorithm $\distalg$ has now the \emph{promise} that in the instances $(G,\inp)$ admissible as inputs, the input vector $\inp$ is restricted to $\inp \in \cI \subset \Sigma^n$.

We define the classes $C_r(t)$ in a way identical to the classes $B_k(t)$, but generalized in two ways. First, the parameter $r$ is not bounded to be integral, but can be any positive real. Second, the decision problems under consideration are extended to the ones in which the set of input vectors $\inp$ can be restricted. So, in particular, $B_k(t) \subseteq C_k(t)$, for every positive integer $k$, and every function $t$. The following theorem proves that boosting can made as limited as desired. The proof of this theorem is deferred to the Appendix.
\begin{theorem}
\label{theo:seperation_rationals}
Let $r < r'$ be any two positive reals. Then, $C_{r'}(0) \setminus C_{r}(t) \neq \emptyset$ for every $t=o(n)$.
\end{theorem}
\def\APPENDRATIONAL{
Let $\widehat{r}=a/b \in [r,r')$ be a positive rational where $a$ and $b$ are two co-prime integers. By the density of the rational numbers, such $\widehat{r}$ is guaranteed to exist.
To establish the theorem, we consider the language $\leadera$ restricted to instances in $\cI$, where
\[
\cI=\{\inp \in \{0,1\}^{*}  : \; \parallel\inp\parallel_{_1}\;\notin  [a+1,a+b-1]\}.
\]
In other words, the promise says that an input either satisfies $\leadera$, or is far from satisfying $\leadera$ (very many selected nodes). We prove that $\leadera\in C_{r'}(0) \setminus C_{r}(t)$ for every $t=o(n)$.

We begin by showing that $\leadera \in C_{r'}(0)$ by considering the following simple randomized algorithm that runs in $0$ time: every node $v$ which is not selected, i.e., such that $\inp(v)=0$,  says \yes\/ with probability $1$; and every node which is selected, i.e., such that $\inp(v)=1$, says \yes\/ with probability $p^{1/a}$, and \no\/ with probability $1-p^{1/a}$. If the graph has $s\leq a$ nodes selected, then all nodes say \yes\/  with probability $p^{s/a}\geq p$, as desired. Else, there are $s \geq a+b$ leaders, (this follows from the promise), and at least one node says \no\/ with probability $1-p^{s/a} \geq 1-p^{(a+b)/a}=1-p^{1+1/\widehat{r}}$. We therefore get a $(p,q)$-decider with $p^{1+1/\widehat{r}}+q\geq 1$, thus $p^{1+1/r'}+q > 1$ as $r' >\widehat{r}$. It therefore follows that $\leadera \in C_{r'}(0)$.

We now consider the harder direction, and prove that $\leadera \notin C_{r}(t)$, for any $t=o(n)$. Since $\widehat{r}\geq r$, it is sufficient to show that $\leadera \notin C_{\widehat{r}}(t)$.
To prove this separation, consider the $\leadera$ problem restricted to the family of $n$-node paths. Fix a function $t=o(n)$, and assume, towards contradiction, that there exists a distributed $(p, q)$-decider $\distalg$ for $\leadera$ that runs in $O(t)$ rounds, with $p^{1+1/\widehat{r}}+q>1$.  Let $\varepsilon \in (0,1)$ be such that $p^{1+1/\widehat{r}+\varepsilon}+q>1$. Let $P$ be an $n$-node path, and let $S\subset P$ be a subpath of $P$.
Let $\delta \in [0,1]$ be a constant satisfying
\begin{equation}
\label{eq:lambda_res_inp}
0 <\delta < p^{1+1/\widehat{r}}\left( 1-p^{\varepsilon} \right) /(a+b-1).
\end{equation}
Consider a positive  instance and a negative instance of $\leadera$, respectively denoted by
\[
I =(P, \inp) \; \mbox{and} \; I'=(P, \inp').
\]
Both instances are defined on the same $n$-node path $P$, where
\[
n \geq (a+b-1)\left(\SecuritySize(\delta)+1\right)+1.
\]
where $\SecuritySize(\delta)=\SecuritySize(\delta,2t+1$), as defined by Eq.~(\ref{eq:defsecuritysize}).
We consider executions of $\distalg$ on these two instances,
where nodes are given the  same id's.
Both instances have almost the same input. In particular,
the only difference is that instance $I$ contains $a$ selected nodes,
whereas $I'$ has the same selected nodes as $I$ plus $b$ additional selected nodes.  Therefore $I$ is legal, while $I'$ is illegal. In addition, both inputs $\inp$ and $\inp'$ satisfy the promise.

In $I'$, the path $P$ is composed of $a+b$ sections, each containing a unique selected node,
and where each pair of consecutive sections separated by $\delta$-secure subpaths.
More precisely, let us enumerate the nodes of $P$ from $1$ to $n$, with node $v$ adjacent to nodes $v-1$ and $v+1$, for every $1<v<n$. Consider the $a+b-1$ subpaths of $P$ defined by:
\[S_i=[(i-1)\SecuritySize(\delta)+i+1,i \cdot \SecuritySize(\delta)+i]\]
for $i=\{1, \ldots, a+b-1\}$.
Let the selected nodes in $I'$ be positioned as follows. Let $u_1=1$ and let $u_i=(i-1)\SecuritySize(\delta)+i$ for $i=2, \ldots, a+b$. Then set
\[
\inp'(v)=\left\{\begin{array}{ll}
1 & \mbox{if $v=u_i$ for some $i \in \{1, ..., a+b\}$}\\
0 & \mbox{otherwise.}
\end{array}\right.
\]
Our next goal is to construct a legal input $I=(P,\inp)$ with $a$ leaders. Towards this, we begin by showing that each $S_i$ contains a $\delta$-secure internal  subpath $S_i'=[x_i,y_i]$ (internal to $S_i$). Note that Fact~\ref{cl:sec_zone_guarantee} refers to subpaths in \emph{valid} instances $(P,\inp) \in \cL$, and since $I'$ is illegal it cannot be directly applied. So instead, let us focus on the instance $(S_i, \inp'_{S_i})$ with IDs $\id_{S_i}$. Since $S_i$ contains no leaders, $\parallel\inp'_{S_i}\parallel_{_1}\;= 0$, it follows that $(S_i, \inp'_{S_i}) \in \cL$. Now we can safely apply Fact~\ref{cl:sec_zone_guarantee}. Indeed, since $|S_i|> \SecuritySize(\delta)$ it follows by the fact that $S_i$ contains an internal $\delta$-secure subpath $S'_{i}=[x_i,y_i]$. Therefore, when applying algorithm $\distalg$ on $(S_i, \inp'_{S_i},\id_{S_i})$ and on $(P, \inp',\id)$, the nodes of $S'_i$ behave the same, thus $\Pr[\cE(P,\inp',\id,V(S'_i))]=\Pr[\cE(S_i,\inp'_{S_i},\id_{S_i},V(S'_i))]$. Hence, $S'_i$ is a $\delta$-secure subpath in $I'$ as well, for every $i \in \{1, ..., a+b-1\}$.

The $\delta$-secure subpaths $S'_i$ are used to divide $P$ into $2(a+b-1)+1$ segments. There are $a+b$ segments $T_{i}$, $i=1,\dots,a+b$, each with one selected nodes. The $\delta$-secure subpaths $S'_i=[x_i,y_i]$ separate $T_{i}$ from $T_{i+1}$. More precisely, we set
\begin{eqnarray*}
T_{1}=[1, x_1-1], \; T_i=[y_{i-1}+1, x_i-1]
\end{eqnarray*}
for $i \in {2, ..., a+b-1}$, and $T_{a+b}=[y_{a+b}+1,n]$, getting
\[
P= T_{1} \circ S'_1 \circ T_2 \circ S'_2 \circ \ldots \circ T_{a+b-1} \circ S'_{a+b-1} \circ T_{a+b}
\]
where $\circ$ denotes path concatenation.
For $i \in \{1, ..., a+b\}$, let $\mathcal{T}_i=\cE(P,\inp',\id,V(T_i))$ be the event that all nodes in the subpath $T_{i}$ say \yes\/ in instance $I'$, and let $p_i=\Pr[\mathcal{T}_i]$ its probability. Let $J=\{j_1, \ldots, j_{b}\}$ be the set of $b$ indices with maximal values in $\{p_1, \ldots, p_{a+b}\}$. I.e.,
$p_{j} \geq \max\{p_i \mid i \in \{1,..., a+b\} \setminus J\}$ for every $j \in J$.
We are now defining the valid instance $I=(P, \inp)$:
\[
\inp(v) = \left\{\begin{array}{ll}
1 &  \mbox{if $v=u_i$ for some $i \in \{1, ..., a+b\} \setminus J$} \\
0 &   \mbox{otherwise.}
\end{array}\right.
\]
We therefore have that $\parallel\inp'\parallel_{_1}\;= a+b$ and $\parallel\inp \parallel_{_1}\;= a$, thus $I \in \leadera$ while $I'\notin \leadera$, and both $I,I'$ satisfy the promise. 
We now make the following immediate observation.
\begin{claim}
$\Pr[\cE(P,\inp,\id,V(T_i))]=p_i$, for every $i \notin J$.
\end{claim}
This follows since the distance between any two nodes $u$ (resp., $v$) in distinct  $T_{i}'s$ is greater than $t$, which implies that $\inp(L_i \circ T_i \circ R_i) = \inp'(L_i \circ T_i \circ R_i)$ where $L_i$ (resp., $R_i$) is the subpath of length $t$ to the left (resp., to the right) of $T_i$ in $P$, from which it follows that under $\distalg$ the nodes of $T_i$ have the same behavior in both instances $I$ and $I'$ for every $i \notin J$.
\QED
Let $\cN$ (resp., $\cN'$) be the event that there exists at least one node in $I$ (resp., $I'$) that says \no\/ when applying algorithm $\distalg$. Similarly, let $\cY$ (resp., $\cY'$) be the event that all nodes in the configuration $I$ (resp., $I'$) say \yes. Let $\mathcal{T}=\bigcup_{i=1}^{a+b} \mathcal{T}_i$ be the event that all nodes in the subpaths $T_{i}$, for $i\in \{1,...,a+b\}$ say \yes\/ in the instance $I'$. For every $i \in \{1, ..., a+b-1\}$, let $\mathcal{S}_i=\cE(P,\inp',\id,V(S'_i))$ be the event that all nodes in the $\delta$-secure subpath $S'_i$ say \yes\/ in the instance $I'$. We have
\begin{eqnarray*}
\Pr(\cY)=\Pr[\cE(P,\inp,\id,V(P))]
\end{eqnarray*}
and
\begin{eqnarray*}
\Pr(\cY')=\Pr[\cE(P,\inp',\id,V(P))]
\end{eqnarray*}
while $$\Pr(\cN)=1-\Pr(\cY)$$ and $$\Pr(\cN')=1-\Pr(\cY').$$
Since $\distalg$ a $(p,q)$-decider, as we assume by contradiction that $\leadera \in B_{k}$, we have $$\Pr(\cN') \geq q,$$ and thus  $$\Pr(\cN') > 1-p^{1+1/\widehat{r}+\varepsilon}.$$ Therefore, $\Pr(\cY') < p^{1+1/\widehat{r}+\varepsilon}$. Moreover,  since $I \in \leadera$, we also have that $\Pr(\cY)\geq p$. Therefore,
\begin{equation}
\label{eqn:ratio_k_upper_res_inp}
\frac{\Pr(\cY')}{\Pr(\cY)} < p^{1/\widehat{r}+\varepsilon}~.
\end{equation}
On the other hand, by applying the union bound to the $a+b$ events $\mathcal{T}, \bigcup_{i=1}^{a+b-1}\mathcal{S}_{i}$, we get that
\begin{eqnarray*}
\Pr(\cN') &\leq& \left(1-\Pr[\mathcal{T}] \right) +\sum_{i=1}^{a+b-1}(1-\Pr[\mathcal{S}_{i}])
\\ &\leq&
1-(\prod_{i \notin J} p_i \cdot \prod_{j \in J} p_j )+(a+b-1) \cdot \delta,
\end{eqnarray*}
where the last inequality follows since $S'_i$'s are $\delta$-secure subpaths and hence the events $\mathcal{T}_i$'s are independent.
We therefore get that
\[
\Pr(\cY') \geq \left(\prod_{i \notin J} p_i \cdot \prod_{j \in J} p_j \right) -(a+b-1) \cdot \delta~.
\]
Since $\Pr(\cY) \leq \prod_{i \notin J} p_i$, it then follows that
\begin{eqnarray*}
\frac{\Pr(\cY')}{\Pr(\cY)} &\geq& \frac{\prod_{i \notin J} p_i \cdot \prod_{j \in J} p_j -(a+b-1) \cdot \delta}{\prod_{i \notin J} p_i}
\\&\geq&
\prod_{j \in J} p_j -\frac{(a+b-1) \cdot \delta}{\prod_{i \notin J} p_i}.
\end{eqnarray*}
Now, since $I \in \leadera$, we have $\prod_{i \notin J} p_i \geq \Pr(\cY)\geq p$, and thus
\begin{equation}
\label{eqn:ratio_k_lower_res_inp}
\frac{\Pr(\cY')}{\Pr(\cY)}  \geq \prod_{j \in J} p_j -\frac{(a+b-1) \cdot \delta}{p}.
\end{equation}
Note that
\begin{eqnarray}
\label{eqn:prod_res_inp}
\prod_{j \in J} p_j\geq p^{1/\widehat{r}}.
\end{eqnarray}
By the definition of $J$, $|J|=b$, and $\prod_{j \in J} p_j\geq p_i^{b}$ for every $i \notin J$. In addition, since there are $a$ indices $i \notin J$, we get that $(\prod_{j \in J} p_j)^{a/b}\geq \prod_{i \notin J} p_i \geq \Pr(\cY)\geq p$. Combining with the definition of $\widehat{r}$, Eq. (\ref{eqn:prod_res_inp}) follows.
Hence, by Eq. (\ref{eqn:ratio_k_lower_res_inp}), we get  $$\Pr(\cY')/\Pr(\cY)\geq p^{1/\widehat{r}}-\frac{(a+b-1) \cdot \delta}{p}.$$
Combining with Eq. (\ref{eqn:ratio_k_upper_res_inp}) we get that
\[
p^{1/\widehat{r}} -(a+b-1) \cdot \delta/p < p^{1/\widehat{r}+\varepsilon}~,
\]
which is in contradiction to the definition of $\delta$ in Eq. (\ref{eq:lambda_res_inp}). We therefore get that $\leadera \notin C_{\widehat{r}}(t)$, and since $r \leq \widehat{r}$, it also holds that
$\leadera \notin C_{r}(t)$ as required. The theorem follows.
\QED
}
Note that Theorem~\ref{theo:seperation_rationals} demonstrates not only the (almost) inability of boosting the probability of success when the inputs to the nodes are restricted to specific kinds, but also the inability of derandomizing, even above the threshold $p^2+q=1$. Indeed, the following is a direct consequence of Theorem~\ref{theo:seperation_rationals}.
\begin{corollary}
\label{lem:non-seperation-rest-input}
For every positive real $r$, there is a decision problem in $C_r(0)$ which cannot be decided \emph{deterministically} in $o(n)$ rounds.
\end{corollary}


\clearpage
\pagenumbering{roman}
\appendix
\setcounter{equation}{0}

\centerline{\large{\bf APPENDIX}}
\numberwithin{equation}{section}

\section{Some proofs}
\label{append:proofs}

\inline Proof of Fact \ref{fact-secure}:
\APPENDSECURE

\inline Proof of Theorem \ref{thm:seperationinfty}:
\APPENDINFINITY

\inline Proof of Claim \ref{claim-path}:
\APPENDPATHTRIPLE

\inline Proof of Claim \ref{claim-extendable}:
\APPENDPATHEXTEND

\inline Proof of Theorem \ref{theo:seperation_rationals}:
\APPENDRATIONAL

\clearpage
\section{Figures}
\label{append:figures}

\APPENDFIGFACT
\APPENDFIGAMOS
\APPENDFIGTREE
\APPENDFIGPATHS
\clearpage
\def\thepage{}

\commful
\commfulend

\commabs

\commabsend

{\small
\begin{thebibliography}{99}

\bibitem{AKY97}
Y.~Afek, S.~Kutten, and M.~Yung.
\newblock The local detection paradigm and its applications to self
  stabilization.
\newblock {\em TCS}, 186:199--230, 1997.

\bibitem{ABI}
N. Alon, L. Babai, and A. Itai.
\newblock  A fast and simple randomized parallel algorithm for
the maximal independent set problem.
\newblock {\em J. Alg.}, 7:567--583, 1986.

\bibitem{Linial01}
A. Amit, N. Linial, J. Matousek, and E. Rozenman.
\newblock Random lifts of graphs.
\newblock  In {\em Proc. 12th SODA},  883--894, 2001.

\bibitem{APV}
B. ~Awerbuch, B. ~Patt-Shamir, and G. ~Varghese.
\newblock Self-Stabilization By Local Checking and Correction.
\newblock  {\em  Proc. FOCS}, 1991, 268-277.

\bibitem{BM09}
L.  Barenboim and M. Elkin.
\newblock Distributed $(\Delta+1)$-coloring in linear (in delta) time.
\newblock {\em Proc. 41st STOC}, 111--120, 2009.

\bibitem{DHKKNPPW}
A. Das Sarma, S. Holzer,  L. Kor, A.  Korman, D. Nanongkai, G. Pandurangan,
D. Peleg and R. Wattenhofer.
\newblock Distributed Verification and Hardness of Distributed Approximation.
\newblock {\em Proc. 43rd STOC}, 2011.

\bibitem{DP12}
D~Dereniowski and A.~Pelc.
\newblock Drawing maps with advice.
\newblock {\em JPDC},72:132-143, 2012.

\bibitem{D74}
E.W.~Dijkstra.
\newblock Self-stabilization in spite of distributed control.
\newblock {\em Comm. ACM}, 17(11), 643--644, 1974.

\bibitem{silent}
S. ~Dolev, M. ~Gouda, and M. ~Schneider.
\newblock Requirements for silent stabilization.
\newblock {\em Acta Informatica}, 36(6), 447-462, 1999.

\bibitem{FGIP07}
P. Fraigniaud, C. Gavoille, D. Ilcinkas and A. Pelc.
\newblock Distributed Computing with Advice: Information Sensitivity
of Graph Coloring.
\newblock  {\em  Proc. 34th ICALP}, 231-242, 2007.

\bibitem{FIP10}
P.~Fraigniaud, D~Ilcinkas, and A.~Pelc.
\newblock Communication algorithms with advice.
\newblock {\em JCSS}, 76:222--232, 2008.

\bibitem{FKL07}
P.~Fraigniaud, A~Korman, and E.~Lebhar.
\newblock Local MST computation with short advice.
\newblock {\em Proc. 19th SPAA}, 154--160, 2007.

\bibitem{FKP11}
P. Fraigniaud, A. Korman, and D. Peleg.
\newblock Local Distributed Decision.
\newblock \emph{Proc. 52nd FOCS}, 708-717, 2011.

\bibitem{FP12}
P. Fraigniaud and A. Pelc.
\newblock Decidability Classes for Mobile Agents Computing.
\newblock {Proc. 10th LATIN}, 2012.

\bibitem{FRT11}
P. Fraigniaud, S. Rajsbaum, and C. Travers.
\newblock Locality and  Checkability  in Wait-free Computing.
\newblock \emph{Proc. 25th DISC}, 2011.

\bibitem{FRT12}
P. Fraigniaud, S. Rajsbaum, and C. Travers.
\newblock Universal Distributed Checkers and Orientation-Detection Tasks.
\newblock Submitted, 2012.

\bibitem{GS11}
M. G\"o\"os and J. Suomela.
\newblock Locally checkable proofs.
\newblock {\em Proc. 30th PODC}, 2011.

\bibitem{KKP11}
L. Kor, A. Korman and D. Peleg.
\newblock  Tight Bounds For Distributed MST Verification.
\newblock {\em Proc. 28th STACS}, 2011.

\bibitem{KK07}
A.~Korman and S.~Kutten.
\newblock Distributed verification of minimum spanning trees.
\newblock {\em Distributed Computing}, 20:253--266, 2007.

\bibitem{KKM11}
A.~Korman,  S.~Kutten, and T. Masuzawa.
\newblock Fast and Compact Self-Stabilizing  Verification, Computation, and Fault Detection of an MST.
\newblock {\em Proc. 30th PODC}, 2011.

\bibitem{KKP10}
A.~Korman, S.~Kutten, and D~Peleg.
\newblock Proof labeling schemes.
\newblock {\em Distributed Computing}, 22:215--233, 2010.

\bibitem{KSV11}
A.~Korman, J.S.~Sereni, and L.~Viennot.
\newblock
Toward More Localized Local Algorithms: Removing Assumptions Concerning Global Knowledge.
\newblock {\em Proc. 30th PODC}, 49-58, 2011.

\bibitem{K09}
F. Kuhn.
\newblock Weak graph colorings: distributed algorithms and applications.
\newblock {\em Proc. 21st SPAA}, 138--144, 2009.

\bibitem{L86}
M. Luby.
\newblock A simple parallel algorithm for the maximal independent set problem.
\newblock {\em SIAM J.  Comput.}, 15:1036--1053, 1986.

\bibitem{N91}
M. Naor.
\newblock A Lower Bound on Probabilistic Algorithms for Distributive
Ring Coloring.
\newblock  {\em SIAM J. Discrete Math.}, 4(3): 409-412 (1991).

\bibitem{NS93}
M.~Naor and L.~Stockmeyer.
\newblock What can be computed locally?
\newblock {\em SIAM J. Comput.} 24(6): 1259-1277 (1995).

\bibitem{PS96}
A. Panconesi and A. Srinivasan.
On the Complexity of Distributed Network Decomposition.
\newblock {\em J. Alg.} 20: 356-374, (1996).

\bibitem{PelB00}
D. Peleg.
\newblock {\em Distributed Computing: A Locality-Sensitive Approach}.
\newblock SIAM, 2000.

\bibitem{SW10}
J. Schneider and R. Wattenhofer.
\newblock  A new technique for distributed symmetry breaking.
\newblock  In \emph{ Proc. 29th PODC}, 257-266, 2010.

\end{thebibliography}
}


\begin{thebibliography}{99}

\bibitem{AKY97}
Y.~Afek, S.~Kutten, and M.~Yung.
\newblock The local detection paradigm and its applications to self
  stabilization.
\newblock {\em Theoretical Computer Science}, 186(1-2):199--230, 1997.

\bibitem{ABI}
N. Alon, L. Babai, and A. Itai.
\newblock  A fast and simple randomized parallel algorithm for
the maximal independent set problem.
\newblock {\em J. Algorithms}, 7(4):567--583, 1986.

\bibitem{Linial01}
A. Amit, N. Linial, J. Matousek, and E. Rozenman.
\newblock Random lifts of graphs.
\newblock  In {\em Proc. 12th ACM-SIAM Symp. on Discrete Algorithms (SODA)},  883--894, 2001.


\bibitem{APV}
B. ~Awerbuch, B. ~Patt-Shamir, and G. ~Varghese.
\newblock Self-Stabilization By Local Checking and Correction.
\newblock  {\em  Proc. IEEE Symp. on the Foundations of Computer
Science (FOCS)}, 1991, 268-277.


\bibitem{BM09}
L.  Barenboim and M. Elkin.
\newblock Distributed $(\Delta+1)$-coloring in linear (in delta) time.
\newblock {\em Proc. 41st ACM Symp. on Theory of computing (STOC)}, 111--120,
2009.



\bibitem{DHKKNPPW}
A. Das Sarma, S. Holzer,  L. Kor, A.  Korman, D. Nanongkai, G. Pandurangan,
D. Peleg and R. Wattenhofer.
\newblock Distributed Verification and Hardness of Distributed Approximation.
\newblock {\em Proc. 43rd ACM Symp. on Theory of Computing (STOC)}, 2011.


\bibitem{DP12}
D~Dereniowski and A.~Pelc.
\newblock Drawing maps with advice.
\newblock {\em Journal of Parallel and Distributed Computing},72:132-143, 2012.

\bibitem{D74}
E.W.~Dijkstra.
\newblock Self-stabilization in spite of distributed control.
\newblock {\em Comm. ACM}, 17(11), 643--644, 1974.

\bibitem{silent}
S. ~Dolev, M. ~Gouda, and M. ~Schneider.
\newblock Requirements for silent stabilization.
\newblock {\em Acta Informatica}, 36(6), 447-462, 1999.




\bibitem{FGIP07}
P. Fraigniaud, C. Gavoille, D. Ilcinkas and A. Pelc.
\newblock Distributed Computing with Advice: Information Sensitivity
of Graph Coloring.
\newblock  {\em  Proc. 34th Colloq. on Automata, Languages and
Programming (ICALP)}, 231-242, 2007.

\bibitem{FIP10}
P.~Fraigniaud, D~Ilcinkas, and A.~Pelc.
\newblock Communication algorithms with advice.
\newblock {\em J. Comput. Syst. Sci.}, 76(3-4):222--232, 2008.

\bibitem{FKL07}
P.~Fraigniaud, A~Korman, and E.~Lebhar.
\newblock Local MST computation with short advice.
\newblock {\em Proc. 19th ACM Symp. on Parallelism in Algorithms
and Architectures (SPAA)}, 154--160, 2007.

\bibitem{FKP11}
P. Fraigniaud, A. Korman, and D. Peleg.
\newblock Local Distributed Decision.
\newblock \emph{Proc. 52nd Annual IEEE Symposium
on Foundations of Computer Science (FOCS)}, 708-717, 2011.

\bibitem{FP12}
P. Fraigniaud and A. Pelc.
\newblock Decidability Classes for Mobile Agents Computing.
\newblock {Proc. LATIN 2012: Theoretical Informatics - 10th Latin American Symposium}, 2012.

\bibitem{FRT11}
P. Fraigniaud, S. Rajsbaum, and C. Travers.
\newblock Locality and  Checkability  in Wait-free Computing.
\newblock \emph{Proc. 25th International Symposium on Distributed Computing (DISC)}, 2011.

\bibitem{FRT12}
P. Fraigniaud, S. Rajsbaum, and C. Travers.
\newblock Universal Distributed Checkers and Orientation-Detection Tasks.
\newblock Submitted, 2012.

\bibitem{GS11}
M. G\"o\"os and J. Suomela.
\newblock Locally checkable proofs.
\newblock {\em Proc. 30th ACM Symp. on Principles of Distributed Computing (PODC)}, 2011.



\bibitem{KKP11}
L. Kor, A. Korman and D. Peleg.
\newblock  Tight Bounds For Distributed MST Verification.
\newblock {\em Proc. 28th Int. Symp. on Theoretical Aspects of
Computer Science (STACS)}, 2011.

\bibitem{KK07}
A.~Korman and S.~Kutten.
\newblock Distributed verification of minimum spanning trees.
\newblock {\em Distributed Computing}, 20:253--266, 2007.

\bibitem{KKM11}
A.~Korman,  S.~Kutten, and T. Masuzawa.
\newblock Fast and Compact Self-Stabilizing  Verification, Computation, and Fault Detection of an MST.
\newblock {\em Proc. 30th ACM Symp. on Principles of Distributed Computing (PODC)}, 2011.

\bibitem{KKP10}
A.~Korman, S.~Kutten, and D~Peleg.
\newblock Proof labeling schemes.
\newblock {\em Distributed Computing}, 22:215--233, 2010.

\bibitem{KSV11}
A.~Korman, J.S.~Sereni, and L.~Viennot.
\newblock
Toward More Localized Local Algorithms: Removing Assumptions Concerning Global Knowledge.
\newblock {\em Proc. 30th ACM Symp. on Principles of Distributed Computing (PODC)}, 49-58, 2011.

\bibitem{K09}
F. Kuhn.
\newblock Weak graph colorings: distributed algorithms and applications.
\newblock {\em Proc. 21st ACM Symp. on Parallel Algorithms and Architectures
(SPAA)}, 138--144, 2009.








\bibitem{L86}
M. Luby.
\newblock A simple parallel algorithm for the maximal independent set problem.
\newblock {\em SIAM J.  Comput.}, 15:1036--1053, 1986.

\bibitem{N91}
M. Naor.
\newblock A Lower Bound on Probabilistic Algorithms for Distributive
Ring Coloring.
\newblock  {\em SIAM J. Discrete Math.}, 4(3): 409-412 (1991).

\bibitem{NS93}
M.~Naor and L.~Stockmeyer.
\newblock What can be computed locally?
\newblock {\em SIAM J. Comput.} 24(6): 1259-1277 (1995).

\bibitem{PS96}
A. Panconesi and A. Srinivasan.
On the Complexity of Distributed Network Decomposition.
\newblock {\em J. Algorithms} 20(2): 356-374, (1996).


\bibitem{PelB00}
D. Peleg.
\newblock {\em Distributed Computing: A Locality-Sensitive Approach}.
\newblock SIAM, 2000.



\bibitem{SW10}
J. Schneider and R. Wattenhofer.
\newblock  A new technique for distributed symmetry breaking.
\newblock  In \emph{ Proc. 29th ACM Symp. on Principles of Distributed Computing
(PODC)}, 257-266, 2010.



\end{thebibliography}
\end{document}